The differences in SAPA Needs by Route, Traffic Volume and after COVID-19

Katsunobu Okamoto[a], Takuji Takemoto[b], Yoshimi Kawamoto[c], Sachiyo Kamimura[d]

[a]Graduate School of Engineering, University of Fukui, Fukui 910-8507
E-mail: kokamoto@u-fukui.ac.jp
[b]Headquarters for Regional Revitalization, University of Fukui, Fukui910-8507,
E-mail: takemoto@u-fukui.ac.jp
[c]Department of Architecture and Civil Engineering, University of Fukui, Fukui 910-8507
E-mail: yoshimi@u-fukui.ac.jp
[d]Headquarters for Regional Revitalization, University of Fukui, Fukui910-8507,
E-mail: sachi-k@u-fukui.ac.jp

**Abstract：** This study aims to identify the differences in SAPA user interest due to each route, traffic volumes and after COVID-19 by the daily feedback from them and to help develop SAPA plans.

Food was the most common opinion. However, for the route, some showed interest in other options. For the traffic volume, the difference of interest was also shown in some heavy traffic areas, On the other hand, the changes in customer needs after the Covid-19 disaster were less changed.

In addition, there were some differences between the tendency of the user's opinions and the drop-in factors of SAPA in the previous studies.

The bias of the sample data and the collection of samples after the covid-19 are the issues in this research. Since the data may represent some part of users and, due to the covid-19, the demographics of users may change drastically in 2020.

## 1. Introduction

Rest areas on expressways (hereinafter referred to as "SAPA") Since 2005, when the Japan Highway Public Corporation was privatized, SAPA has been improving its services by renovating food courts and shopping corners, restrooms, shower rooms, dog runs, etc., to provide a comfortable environment for customers using the expressway.

There are currently 181 SAPAs[1] (as of April 2020) in the jurisdiction of Central Nippon

---

[1] *NEXCO Central Japan Report* 2020 5p

Expressway Company Limited (hereinafter referred to as "C-NEXCO"). SAPAs are required by the government to be located at standard intervals of 15 km between facilities and standard intervals of 50 km between SAPA, considering the traffic volume and road structure of the main lines.[2] SAPA's basic facilities include parking lots, restrooms, commercial facilities (food courts, shops), and parks, with convenience stores in some locations.

The purpose of this study is to analyze the daily feedback from users to develop a SAPA that better meets the needs of users and to make recommendations on how to reflect the feedback in the SAPA development plan. In addition, it is assumed that users' needs for SAPA are changing due to the recent Covid-19 disaster. In this regard, it is also necessary to analyze the daily opinions.

## 2. Research methods and hypotheses

The composition of vehicles used on expressways differs depending on the route. For example, referring to the average daily traffic volume of all expressways[3], 26.4% of all expressways have medium-sized vehicles or larger, while 31.2% of Tomei Expressway has medium-sized vehicles or larger.

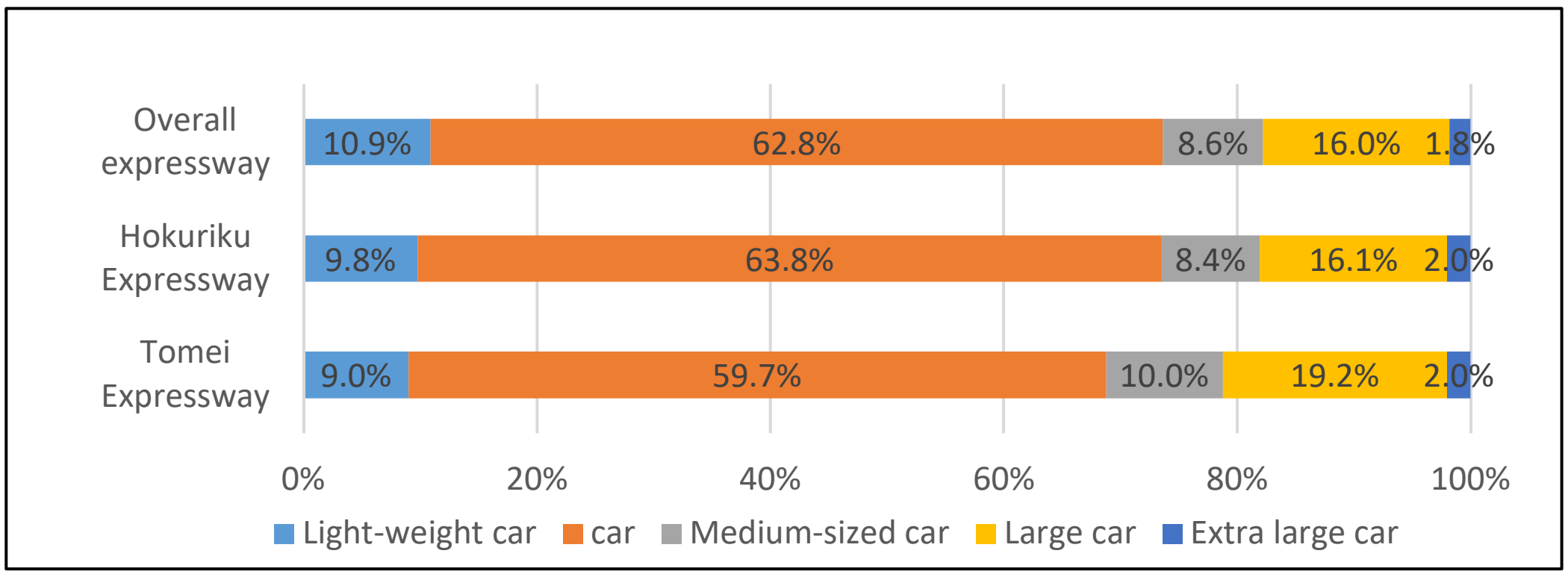

【Figure 1: Average Daily Traffic Volume (FY2015)】

In addition, under the heavy traffic routes, the traffic congestion is more frequent and the driving time is longer compared to other sections. There is a possibility that the need for "rest" will increase even more. Furthermore, considering the impact of covid-19 as the most recent change in the commercial environment, most commercial facilities are required to prevent the bad effect of covid-19 by customers. The hypotheses are below.

H1. If the user composition of the routes is different, the needs for SAPA on each route will also be different.

H2. The traffic volume on the routes connected to SAPA also makes a difference in the needs of the users.

[2] *Expressway Facilitation Committee* 2020 576p
[3] *Expressway Facilitation Committee* 2020 229p

H3. There is a change in the need for SAPA after the covid-19 disaster.

The data to be used for these verifications is the opinions from SAPA users, which are composed of customer calls, e-mails, forms deposited in the customer feedback box, and direct conversations with staff. The period covered was from FY2018 to FY2020, to understand the recent trend and to see the situation before and after the Covid-19 disaster.

In the analysis, text mining methods using KHCoder[45] are used. First, for the classification of interest status, after checking the status of frequently occurring words such as nouns and adjectives in each year's data, reference to hierarchical cluster analysis, and the contents of individual inquiries, I will divide them into the categories of "food" "customer service" "cleanliness" "parking lots" "facility equipment (free)" and "facility equipment (function)". The words targeted by each group are as follows.

| Category | Target words |
|---|---|
| Food | Food name (ramen, curry, etc.), Delicious, Eating, Bad, etc. |
| Customer service | Staff, Shopkeeper, Cashier, Smile, Customer service, Attitude, etc. |
| Cleanliness | Toilet, Sweeping, Cleaning, Cleanliness, Disinfection, Smell, Hand washing, etc. |
| Parking lots | Roadside, Inflow, Guidance, Nighttime, Full, Passenger cars, etc. |
| Facility equipment (free) | Store, Dog park, Children, Private Room, View, park, etc. |
| Facility equipment (function) | Battery Charging, Smoking, Smart, Shower, Gas station, etc. |

【Table 1: Examples of Words Covered by Each Category】

In addition to the above classification, a general classification is added based on "functional behavior" and "free behavior," which are the classifications used by Umayahara et al. (2017)[6]. Those that fall under "functional behavior" are "Food" "parking lots" and" Facility equipment (function)". On the other hand, those corresponding to "free behavior" shall be "customer service" and " Facility equipment (free) ". Cleanliness is treated as a single item because it relates to both behaviors and is difficult to classify. Each hypothesis will be verified using these

[4] Higuchi, k.,2016, "A Two-Step Approach to Quantitative Content Analysis: KH Coder Tutorial Using Anne of Green Gables (Part I )," *Ritsumeikan Social Sciences Review,* 52(3): 77-91

[5] Higuchi, k.,2017, "A Two-Step Approach to Quantitative Content Analysis: KH Coder Tutorial Using Anne of Green Gables (Part2)," *Ritsumeikan Social Sciences Review,* 53(1): 137-47

[6] Atsushi Umayahara, Tetsuya Akagi and Hiroki Suzuki (2017) . Study on the State of Rest Behaviors at Expressway Rest Areas -For Elderly and Families with Children- *J.Archit, Plann., AIJ* 82, P1640)

categories.

The status of the sample data for each year is as follows.

| **Male** | FY2018 | FY2019 | FY2020 |
|---|---|---|---|
| Under 10s | 643 | 873 | 136 |
| 10s | 737 | 1067 | 164 |
| 20s | 534 | 737 | 226 |
| 30s | 573 | 636 | 245 |
| 40s | 1172 | 1190 | 436 |
| 50s | 1358 | 1337 | 619 |
| 60s | 1846 | 1247 | 369 |
| Over 70s | 115 | 693 | 218 |
| Not entered | 1053 | 1581 | 1622 |
| Total | 8031 | 9361 | 4035 |

| **Female** | FY2018 | FY2019 | FY2020 |
|---|---|---|---|
| Under 10s | 1527 | 1378 | 222 |
| 10s | 1202 | 1781 | 254 |
| 20s | 625 | 647 | 211 |
| 30s | 648 | 729 | 198 |
| 40s | 959 | 1059 | 287 |
| 50s | 1062 | 1029 | 302 |
| 60s | 1032 | 748 | 176 |
| Over 70s | 41 | 354 | 101 |
| Not entered | 482 | 531 | 445 |
| Total | 7578 | 8256 | 2196 |

| **Not entered** | FY2018 | FY2019 | FY2020 |
|---|---|---|---|
| Under 10s | 20 | 10 | 0 |
| 10s | 23 | 21 | 3 |
| 20s | 10 | 25 | 3 |
| 30s | 35 | 16 | 9 |
| 40s | 25 | 45 | 6 |
| 50s | 32 | 49 | 15 |
| 60s | 40 | 83 | 15 |
| Over 70s | 2 | 55 | 5 |
| Not entered | 10862 | 2017 | 881 |
| Total | 11049 | 2321 | 937 |
| Sum total | 26658 | 19938 | 7168 |

【Table 2: Sample Data Summary】

From FY2018 to FY2019, the proportions of males and females are almost equal. In terms of age, most males were in 40s to 60s. The total number of cases decreased significantly in FY2020. In addition, the ratio of males is doubled and the number of teenagers has decreased significantly. As for females, teens and under 10 years old are mainly users in FY2018 and FY2019. But, same as mele, it decreased significantly in FY2020, and 30s to 50s became the largest group.

For the first hypothesis, we will classify customer opinions by route and confirm the status of frequent words by attribute and the status of sample data by route using a co-occurrence network for each year, and clarify the differences.

As for the second hypothesis, we will classify the sample data into SAPAs located in areas with heavy traffic and SAPAs located in other areas, confirm the status of frequent words by attribute based on the classification, and analyze the trend of opinions using the multidimensional scaling method to clarify the differences between the two.

For the third hypothesis, the results of the first and second hypothesis analyses will be compared between FY2018, FY2019, and FY2020 to see if there are any changes by route or by traffic volume.

## 3. Pilot study

As for previous studies on the needs of users for SAPA services and facilities, the first one is the study by Umayahara et al. (2017), which investigated the local trends of users in SAPA. In this study, the users of SAPA took functional behaviors corresponding to physiological phenomena, such as using the restroom and eating, as well as free behaviors to enjoy the place, such as using the playground and shopping. While most users are performing functional activities in SAPA, they are also performing free activities as well. Roughly half of their rest time is spent on free activities. It was also found that the maintenance of the park and the playground was an incentive factor.

In addition, Nishii et al. (2017)[7] found that there are two classes of SAPAs users. Class 1: they use expressways for homecoming and truck bus users. Class 2: they use expressways for sightseeing and vacation and passenger cars. They also found that the items that affect the use of SAPA in each class were clarified. In Class 1, the driver's level of fatigue was the basis of the decision to stop at SAPA, and there was a tendency to use places far from the expressway entrance. On the other hand, Class 2 tends to stop at SAPA as a destination regardless of the distance from the expressway entrance. Shiino et al. (2011)[8] found that driver's condition, SAPA's facilities (such as convenience stores), and traffic congestion influence increasing the rate of stopping at SAPA.

According to Nishii et al (2017), the needs of SAPAs users are different by vehicle type. So I think that the differences in the composition of driven on each route may lead to differences in the needs of each route. In addition, Shiino et al. (2011) show that the level of fatigue and traffic congestion affects SAPA drop-offs. It suggests that differences in traffic volume on connecting routes affect the need for SAPA.

## 4. Users' needs by route

To identify users' needs by route, we targeted those routes whose names were registered in the "Route" category in the sample data. The number of data by route is as follows.

---

[7] Kazuo Nishii, Hideki Furuya and Kuniaki Sasaki (2017). Market Segmentation Modeling of Tourism & Leisure Vehicle Driver's Behaviors of Stopover at SA/PAs on Expressway - Through application of Latent Class Analysis to Chugoku Expressway data - *Journal of the City Planning Institute of Japan*, 52, 1266p)

[8] Osamu Shiino, Naohiko Hibino, Shigeru Morichi (2011). Stopping Characteristics and Congestion Countermeasures for Highway Rest Facilities 4p~7p

| Road name | FY2018 | FY2019 | FY2020 |
|---|---|---|---|
| Aboutouge road | 1 | 0 | 0 |
| Ise Expressway | 454 | 229 | 69 |
| Ise Bay Coastal Expressway | 131 | 93 | 79 |
| Kisei Expressway | 21 | 18 | 10 |
| Ken-O Expressway | 420 | 397 | 120 |
| Odawara atsugi road | 134 | 147 | 73 |
| Shin-Tomei Expressway | 6113 | 4380 | 1166 |
| New Meishin Expressway | 149 | 313 | 100 |
| Seisho Bypass | 65 | 39 | 39 |
| Chuo Expressway | 4933 | 3134 | 1005 |
| Central Transect Expressway | 1 | 12 | 5 |
| Nagano Expressway | 283 | 190 | 90 |
| Tokai Kanjo Expressway | 63 | 52 | 59 |
| Tokai-Hokuriku Expressway | 346 | 319 | 226 |
| East Fuji Five Lakes Road | 0 | 2 | 0 |
| Tomei Expressway | 7759 | 5846 | 2414 |
| Higashimeihan Expressway | 718 | 420 | 174 |
| Maizuru-Wakasa Expressway | 21 | 9 | 2 |
| Hokuriku Expressway | 2535 | 2056 | 639 |
| Meishin Expressway | 1948 | 1627 | 514 |
| Nagoya second ring road | 0 | 3 | 3 |
| Total | 26658 | 19938 | 7168 |

【Table 3: Number of Samples by Route】

The routes with a sample size of less than 70 in some years are excluded from the verification because the analysis of the results cannot be done sufficiently (the routes with gray squares in the above table are included).

The following table shows the classification of the routes by interest category for each year.

| | FY2018 | | FY2019 | | FY2020 | |
|---|---|---|---|---|---|---|
| | Number | Percentage | Number | Percentage | Number | Percentage |
| Food | 29,993 | 35.27% | 11,011 | 55.22% | 3,115 | 43.38% |
| Customer service | 4,833 | 5.68% | 3,176 | 15.93% | 1,288 | 17.94% |
| Cleanliness | 6,181 | 7.27% | 4,502 | 22.58% | 2,084 | 29.03% |
| Parking lots | 6,663 | 7.84% | 4,173 | 20.93% | 2,199 | 30.63% |
| Facility equipment (free) | 1,882 | 2.21% | 1,110 | 5.57% | 718 | 10.00% |
| Facility equipment (function) | 6,045 | 7.11% | 3,516 | 17.63% | 1,656 | 23.06% |

【Table 4: Interest Category Classification by Fiscal Year】

Based on this sample data, the "Interest Category Status by Route," which shows the frequency and percentage of each category received for each route, and the "Interest Category Co-occurrence Network Diagram," which shows the relationship between the route name and each category, will be used to clarify the trends by the route.

## 4-1. Status for each year

### 4-1-1. FY 2018

The interest category status by route for FY2018 is as follows.

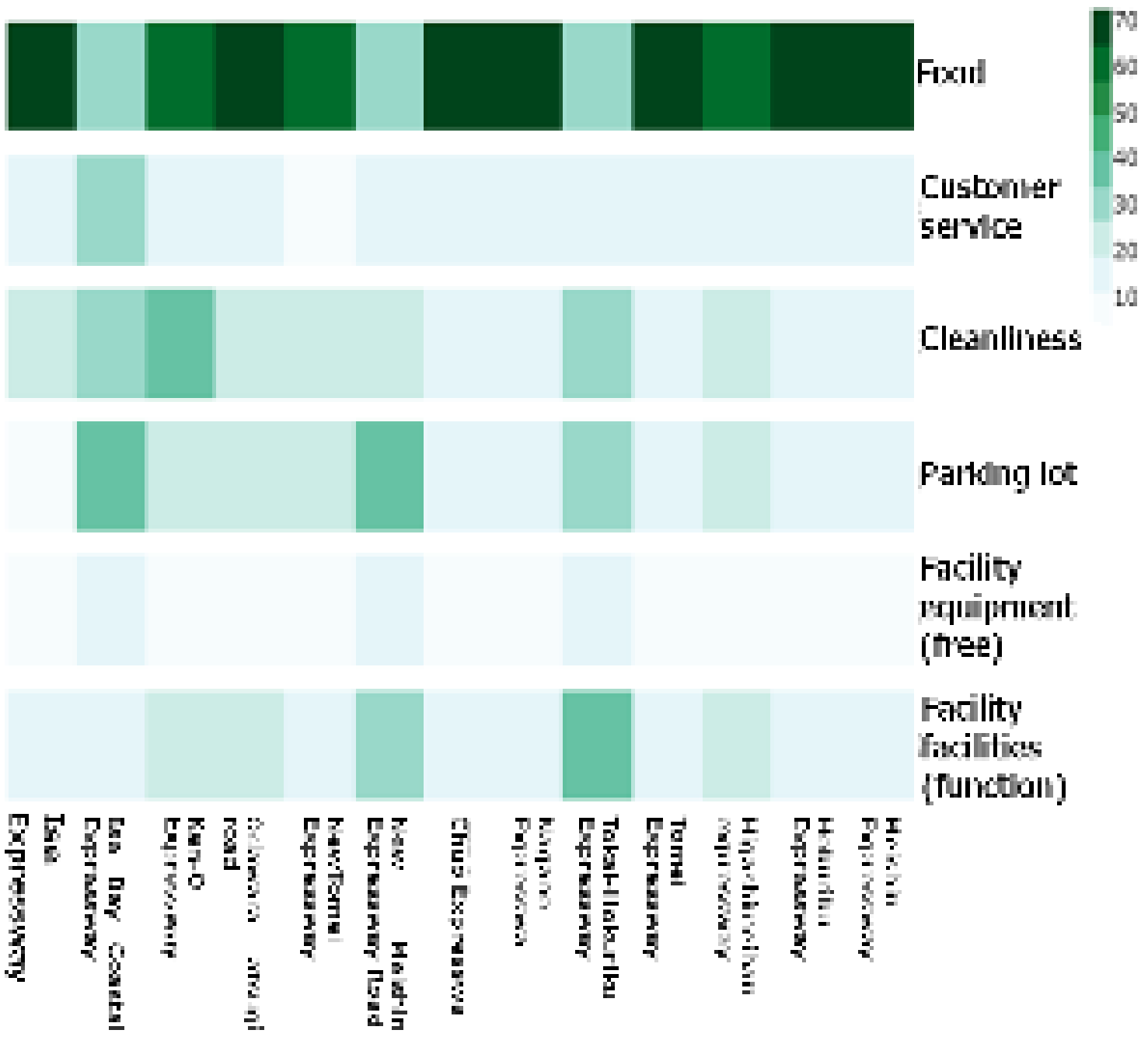

【Figure 2: Interest Category Heat Chart (FY2018)】

Most of the routes were interested in food. While Ise Wangan expressway and New Meishin expressway were most interested in parking lots. Tokai-Hokuriku expressway was most interested in facility facilities (free). The Ise Wangan expressway and Shin Meishin expressway were most interested in parking lots.

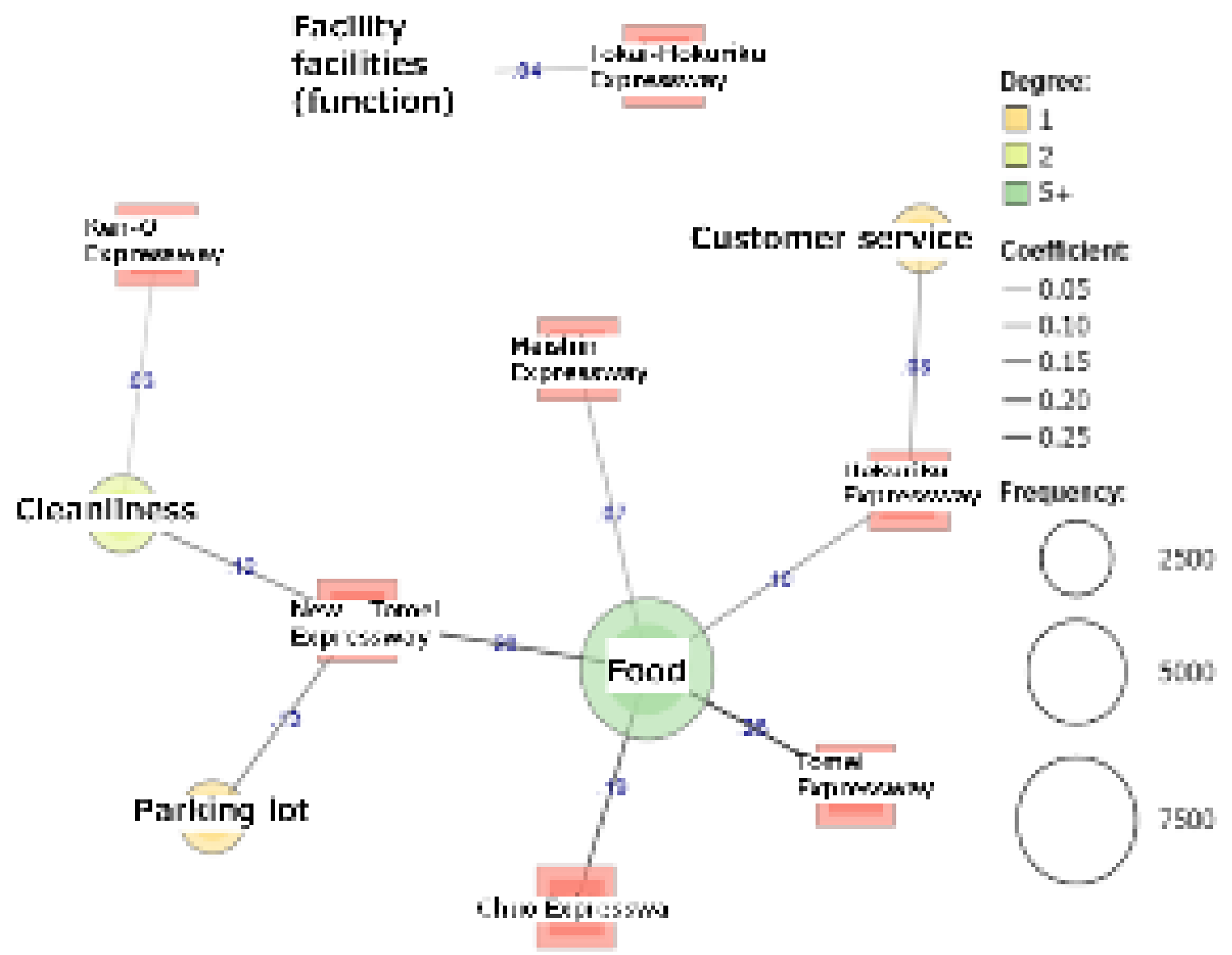

(Settings)
Coding unit ：H5
Selecting co-occurrence relations：
Select the top 10 calculated by Jaccard coefficient.
Factor ： Standardization

【Figure 3: Interest Category Co-occurrence Network Diagram (FY2018)】

The status of the co-occurrence network diagram is shown on the left. The Tomei Expressway and the New Tomei expressway have a strong correlation of 0.2 or more with meals, and the Chuo expressway and the Hokuriku expressway correlate 0.1. For the New Tomei expressway, there is also a correlation with parking lots (0.13) and cleanliness (0.12).

These results can be summarized as follows.

| User Interest Trends | Road name |
|---|---|
| Food | Ise expressway, Ken-O expressway, Odawara atsugi road, New Tomei expressway, Chuo expressway, Nagano expressway, Tomei expressway, Higashimeihan expressway. Hokuriku expressway, Meishin expressway |
| Cleanliness | New Tomei expressway |
| Parking lots | New Tomei expressway, New Meishin expressway, Ise Bay Coastal expressway |
| Facility equipment (free) | Tokai-Hokuriku expressway |

【Table 5: Trends in User Interest by Route (FY2018)】

## 4-1-2. FY 2019

The aggregate results for 2019 are as follows.

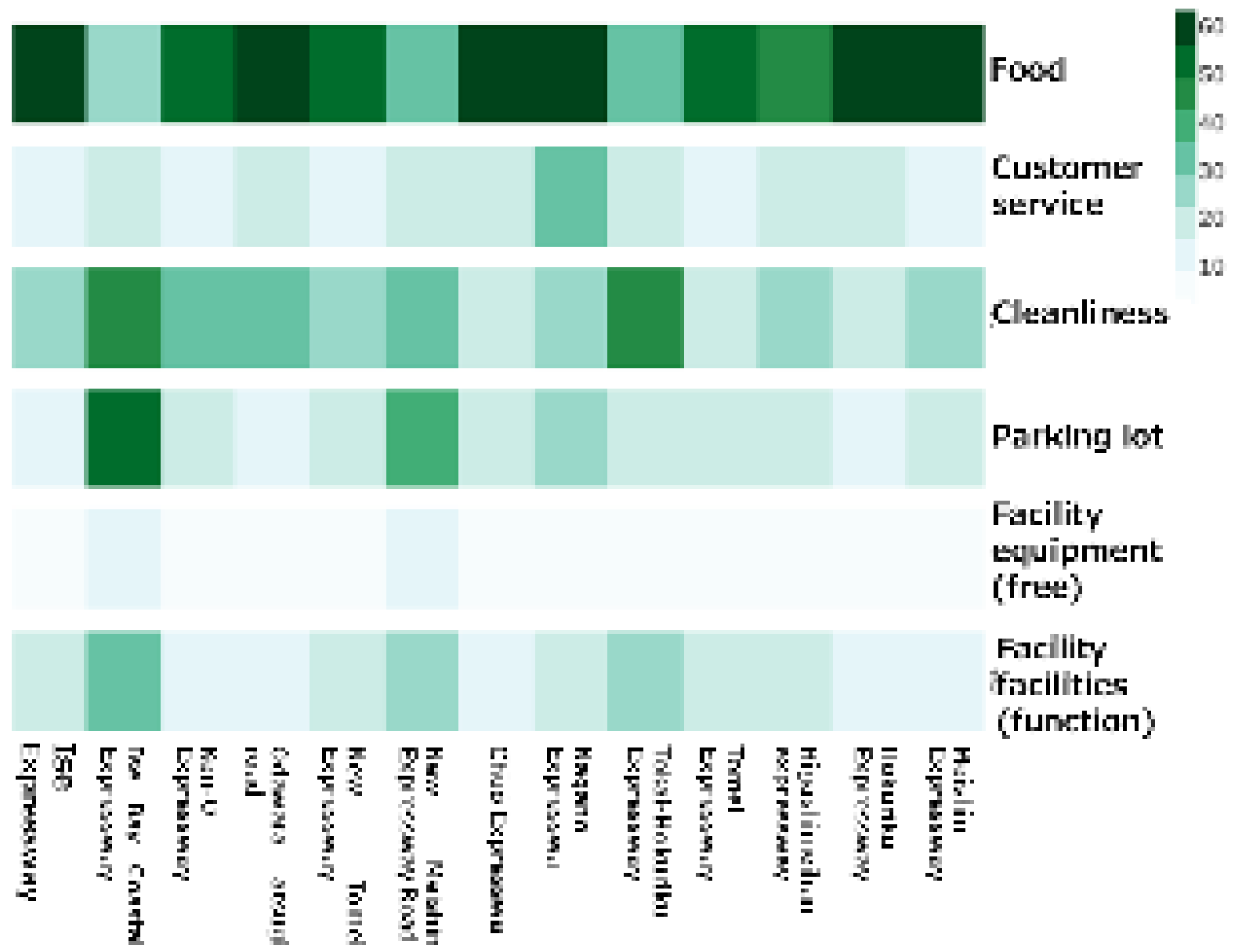

【Figure 4: Interest Category Heat Chart (FY2019)】

Most routes with the highest level of interest in food were the same as in the previous year. While parking lots were the most important factor for the Ise Wangan expressway and Shin Meishin expressway. Cleanliness was the most important factor for the Tokai-Hokuriku expressway.

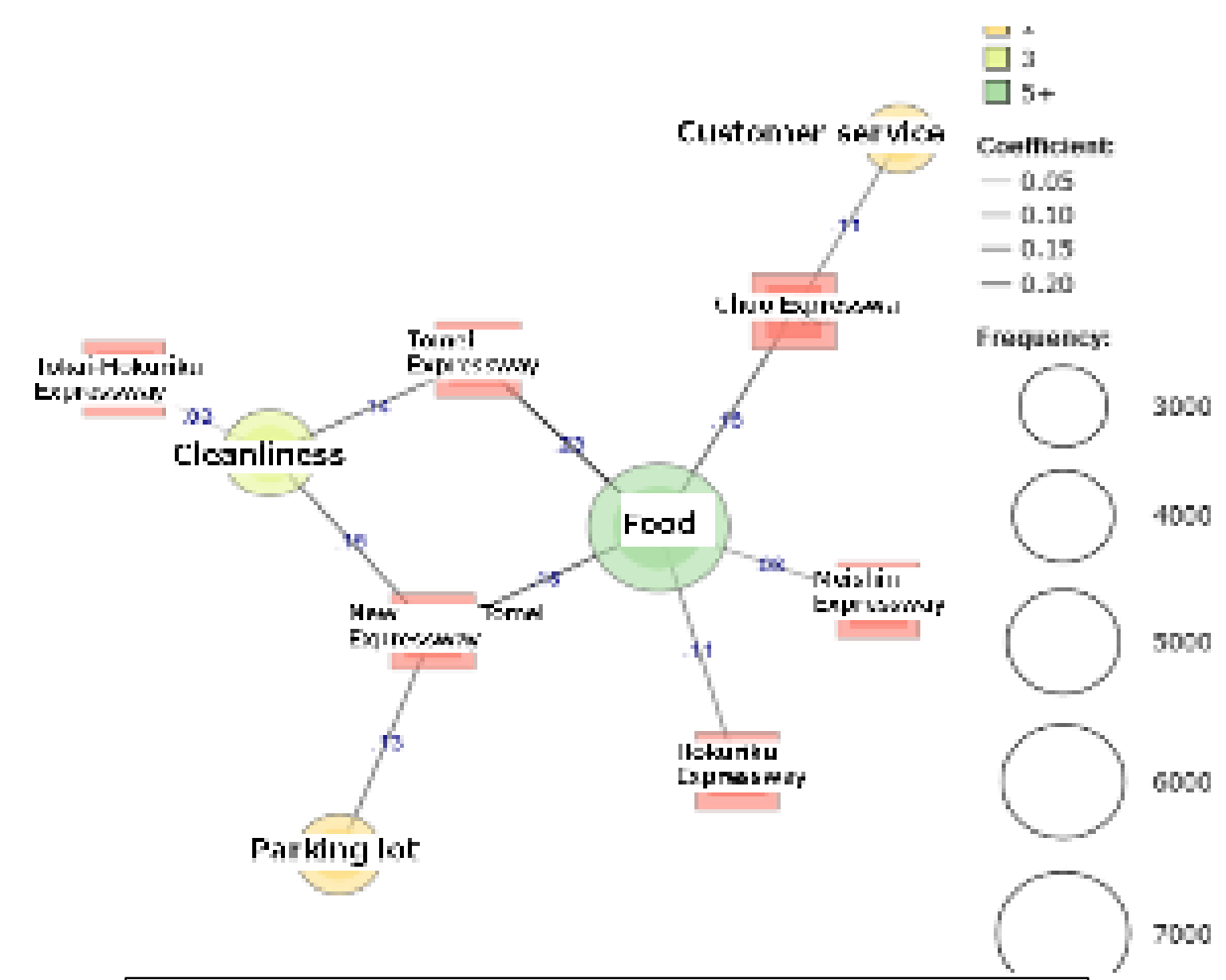

(Settings)
Coding unit ：H5
Selecting co-occurrence relations：
Select the top 10 calculated by Jaccard coefficient.
Factor ： Standardization

【Figure 5: Interest Category Co-occurrence Network Diagram (FY2019)】

The co-occurrence network diagram is as follows.

As for food, Tomei expressway, New Tomei expressway, and Chuo expressway show strong correlations of 0.23, 0.18, and 0.16, respectively, followed by Hokuriku expressway with 0.11 and Meishin expressway with 0.08. As for the New Tomei Expressway, there is a correlation of 0.15 with cleanliness. A correlation of 0.11 was also found between the Central expressway and customer service.

These can be summarized as follows.

| User Interest Trends | Road name |
|---|---|
| Food | Ise Expressway, Ken-O Expressway, Odawara atsugi road, New Tomei Expressway, Chuo Expressway, Nagano Expressway, Tomei Expressway, Higashimeihan Expressway .Hokuriku Expressway, Meishin Expressway |
| Customer service | Chuo Expressway |
| Cleanliness | New Tomei Expressway, Tomei Expressway, Tokai-Hokuriku Expressway |
| Parking lots | New Meishin Expressway, Ise Bay Coastal Expressway |

【Table 6: Trends in User Interest by Route (FY2019)】

### 4-1-3. FY 2020

The aggregate results for FY2020 are as follows.

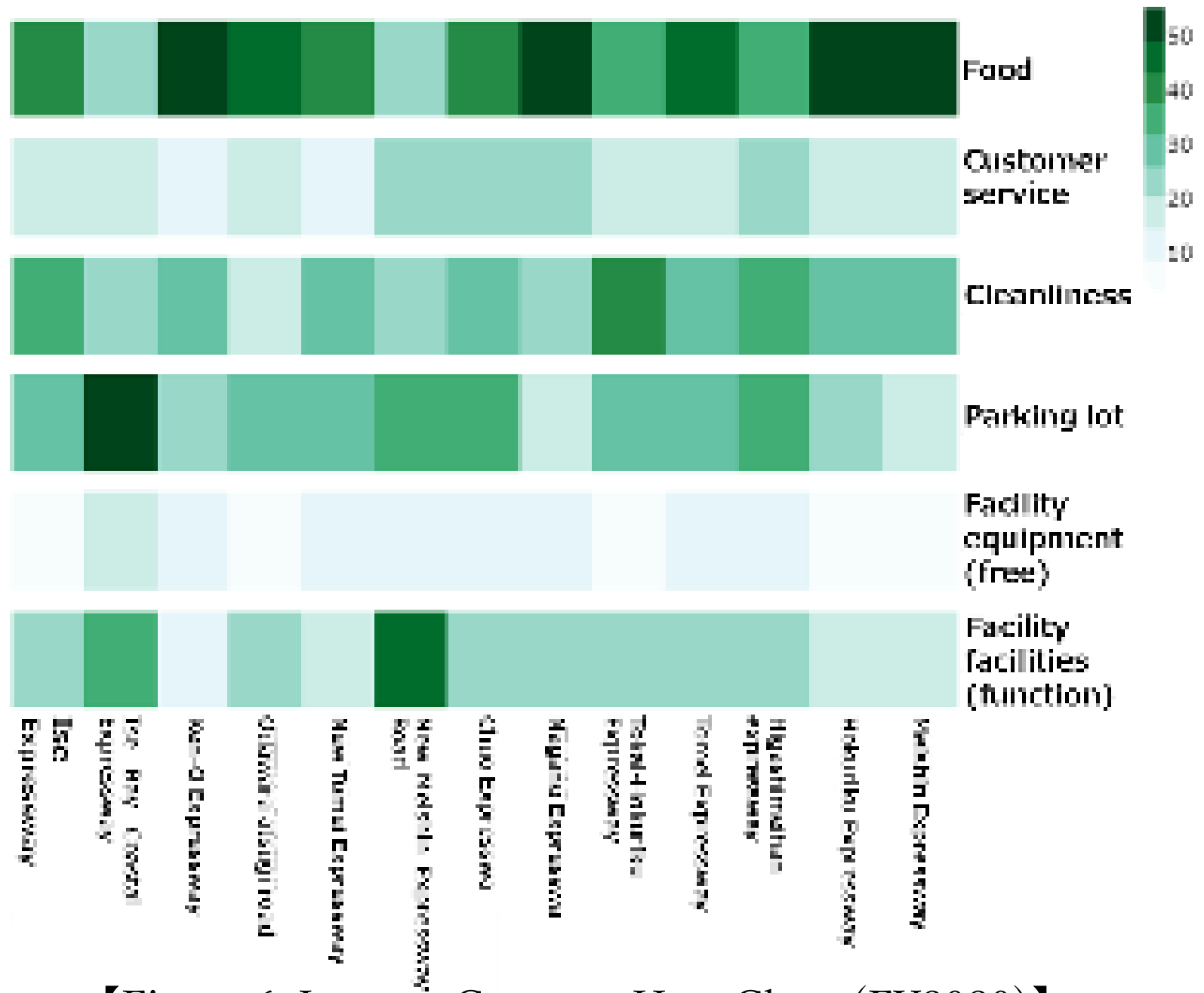

【Figure 6: Interest Category Heat Chart (FY2020)】

As in the previous year, 10 routes showed the highest level of interest in food. While the Ise Bay Coastal expressway ranked parking lots, the Shin Meishin expressway ranked facility equipment (free), and the Tokai-Hokuriku expressway ranked cleanliness.

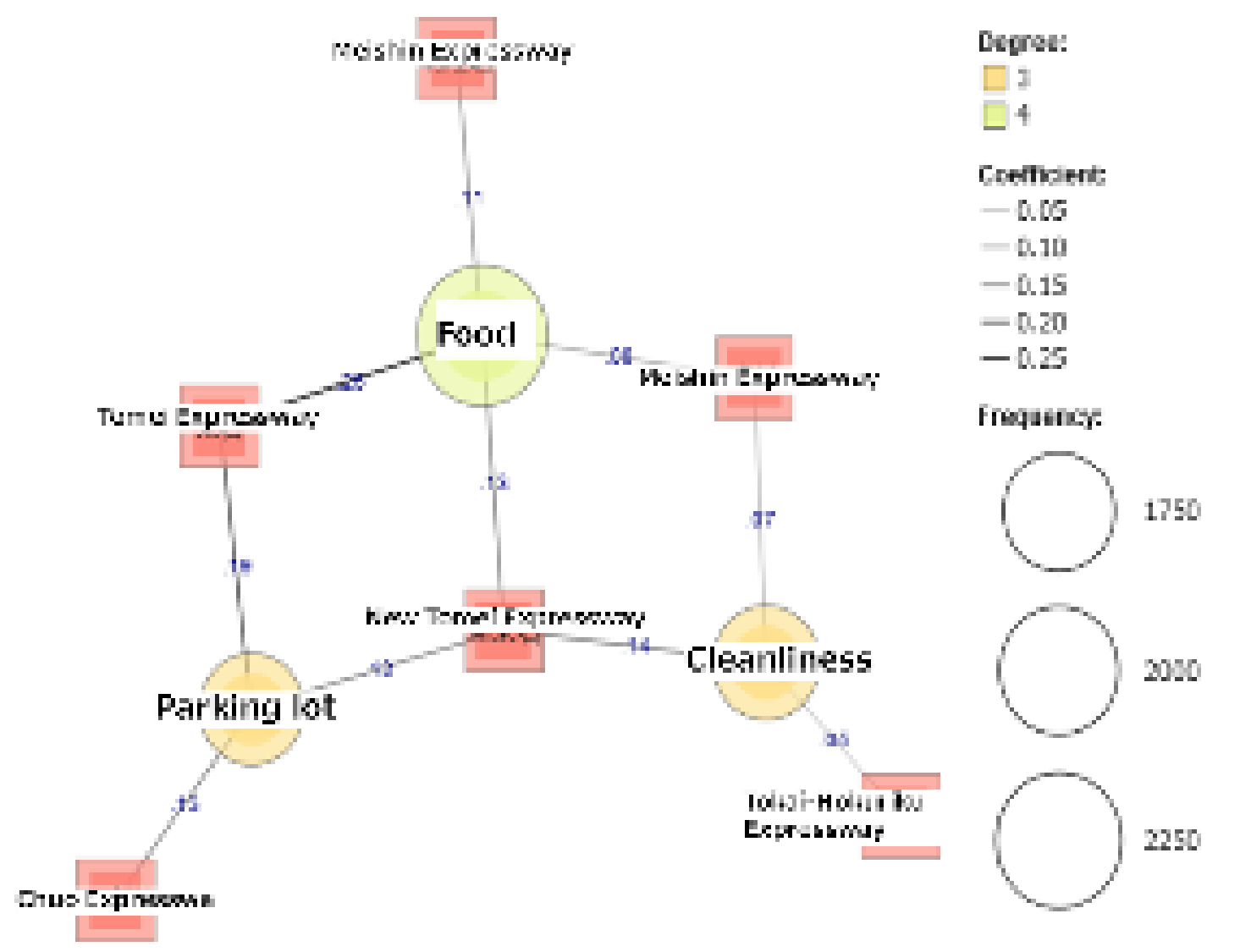

(Settings)
Coding unit ：H5
Selecting co-occurrence relations：
Select the top 10 calculated by Jaccard coefficient.
Factor ： Standardization

【Figure 7: Interest Category Co-occurrence Network Diagram (FY2020)】

The co-occurrence network diagram is shown on the left. The Tomei expressway showed a strong correlation with meals at 0.2, the New Tomei expressway at 0.13, the Hokuriku expressway at 0.11, and the Meishin expressway at 0.08. The Chuo expressway showed a correlation of 0.13 with parking lots. The New Tomei expressway also correlated with this item at 0.13. On the other hand, the correlation of cleanliness was 0.14 with the New Tomei expressway and 0.07 with the Meishin expressway.

These can be summarized as follows.

| User Interest Trends | Road name |
|---|---|
| Food | Ise Expressway,Ken-O Expressway, Odawara atsugi road, New Tomei Expressway, Chuo Expressway, Nagano Expressway, Tomei Expressway, Higashimeihan Expressway. Hokuriku Expressway, Meishin Expressway |
| Cleanliness | New Tomei Expressway, Tokai-Hokuriku Expressway |
| Parking lots | Ise Bay Coastal Expressway, New Tomei Expressway, Tomei Expressway, Chuo Expressway |
| Facility equipment (free) | New Meishin Expressway |

【Table 7: Trends in User Interest by Route (FY2020)】

### 4-2. The characteristics of each route

The relationship between the user interest categories and each route for each year is shown below.

| User Interest Trends | FY2018 | FY2019 | FY2020 |
|---|---|---|---|
| Food | Ise Expressway, Ken-O Expressway, Odawara atsugi road, New Tomei Expressway, Chuo Expressway Nagano Expressway, Tomei Expressway, Higashimeihan Expressway. Hokuriku Expressway, Meishin Expressway | Ise Expressway, Ken-O Expressway, Odawara atsugi road, New Tomei Expressway, Chuo Expressway Nagano Expressway, Tomei Expressway, Higashimeihan Expressway. Hokuriku Expressway, Meishin Expressway | Ise Expressway, Ken-O Expressway, Odawara atsugi road, New Tomei Expressway, Chuo Expressway Nagano Expressway, Tomei Expressway, Higashimeihan Expressway, Hokuriku Expressway, Meishin Expressway |
| Customer service | ― | Chuo Expressway | ― |
| Cleanliness | New Tomei Expressway | New Tomei Expressway, Tomei Expressway, Tokai-Hokuriku Expressway | New Tomei Expressway, Tokai-Hokuriku Expressway |
| Parking lots | New Tomei Expressway, New Meishin Expressway, Ise Bay Coastal Expressway | New Meishin Expressway, Ise Bay Coastal Expressway | Ise Bay Coastal Expressway, New Tomei Expressway, Tomei Expressway, Chuo Expressway |
| Facility equipment (free) | Tokai-Hokuriku Expressway | _ | New Meishin Expressway |

【Table 8: Trends in User Interest by Route in All Periods】

Throughout the period under study, interest in food was high on many routes. Despite this trend, there were some routes where people were interested in something other than food.

No major changes were seen in terms of pre and post Covid-19, and in FY2020, there was still a high level of interest in food on many routes. There was no increase in the number of routes with high interest in cleanliness, which is assumed to be directly related to Covid-19.

## 5. Differences in customer needs due to traffic volume

In order to clarify whether the needs change depending on the traffic volume on the routes connected to SAPA, the traffic volume is classified into heavy traffic sections when the cross-sectional traffic volume of the section connected to SAPA is 50,000 vehicles or more on average per day, and non-heavy traffic sections for other sections. This figure of 50,000 vehicles is a rough estimate of the number of vehicles that will be congested due to traffic concentration, and it is assumed that there will be changes in the purpose of using SAPA, such as avoiding traffic congestion and reducing driving fatigue due to traffic concentration.

Based on the "Changes in Traffic Volume on Major Sections of Expressways and Major National Highways in Japan and Major Urban Areas"[9] published by the Ministry of Land, Infrastructure, Transport and Tourism, the relevant sections are shown in the figure below.

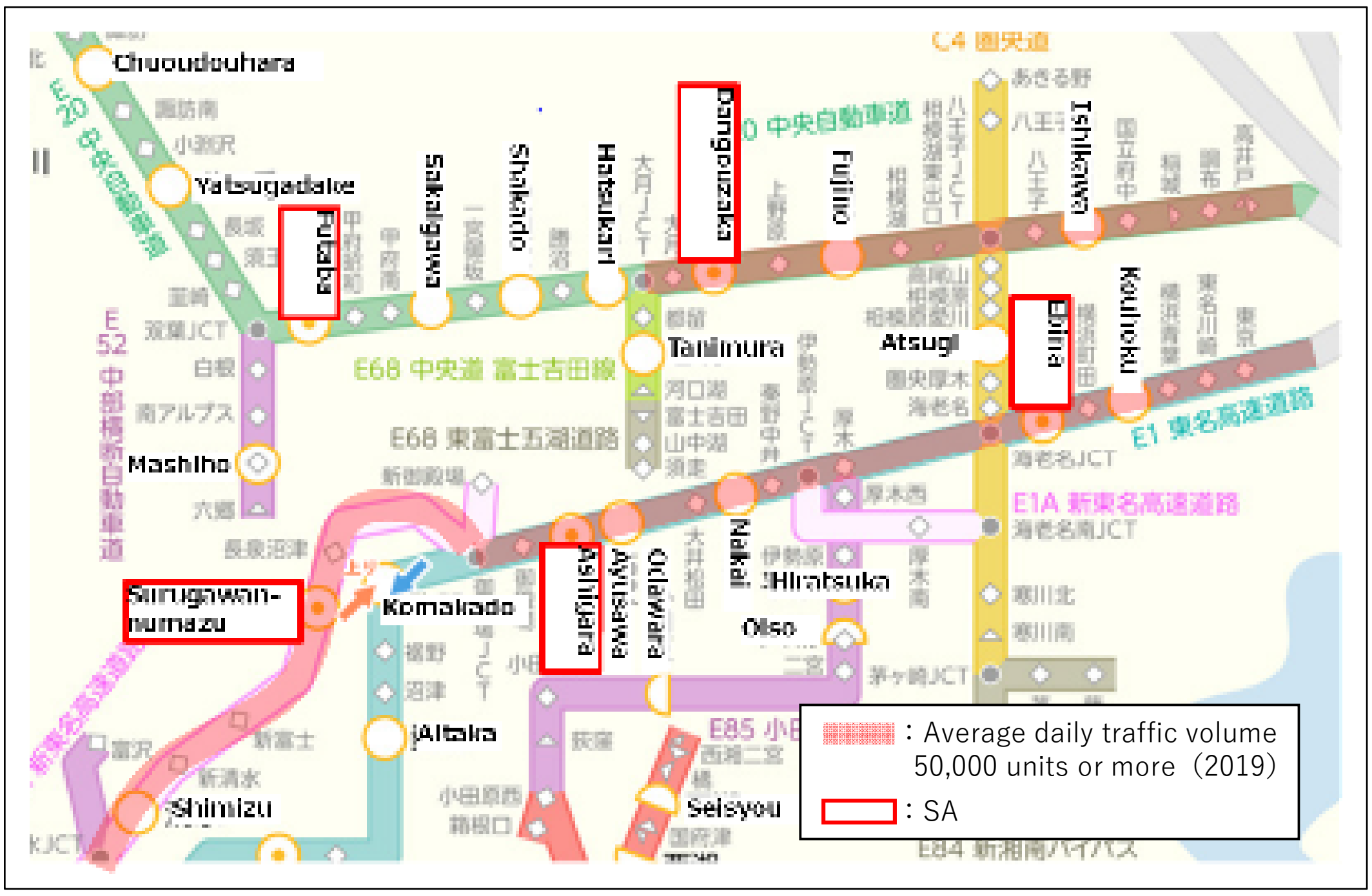

【Figure 8: Heavy Traffic Sections in the Tokyo Metropolitan Area[10]】

---

9 Changes in Traffic Volume on Major Sections of Expressways and Major National Highways in Japan and Major Urban Areas (FY2019). Ministry of Land, Infrastructure, Transport and Tourism (MLIT) website, (https://www.mlit.go.jp/road/road_fr4_000090.html)

10 Central Nippon Expressway Company Limited HP (https://sapa.c-nexco.co.jp/search/highway).

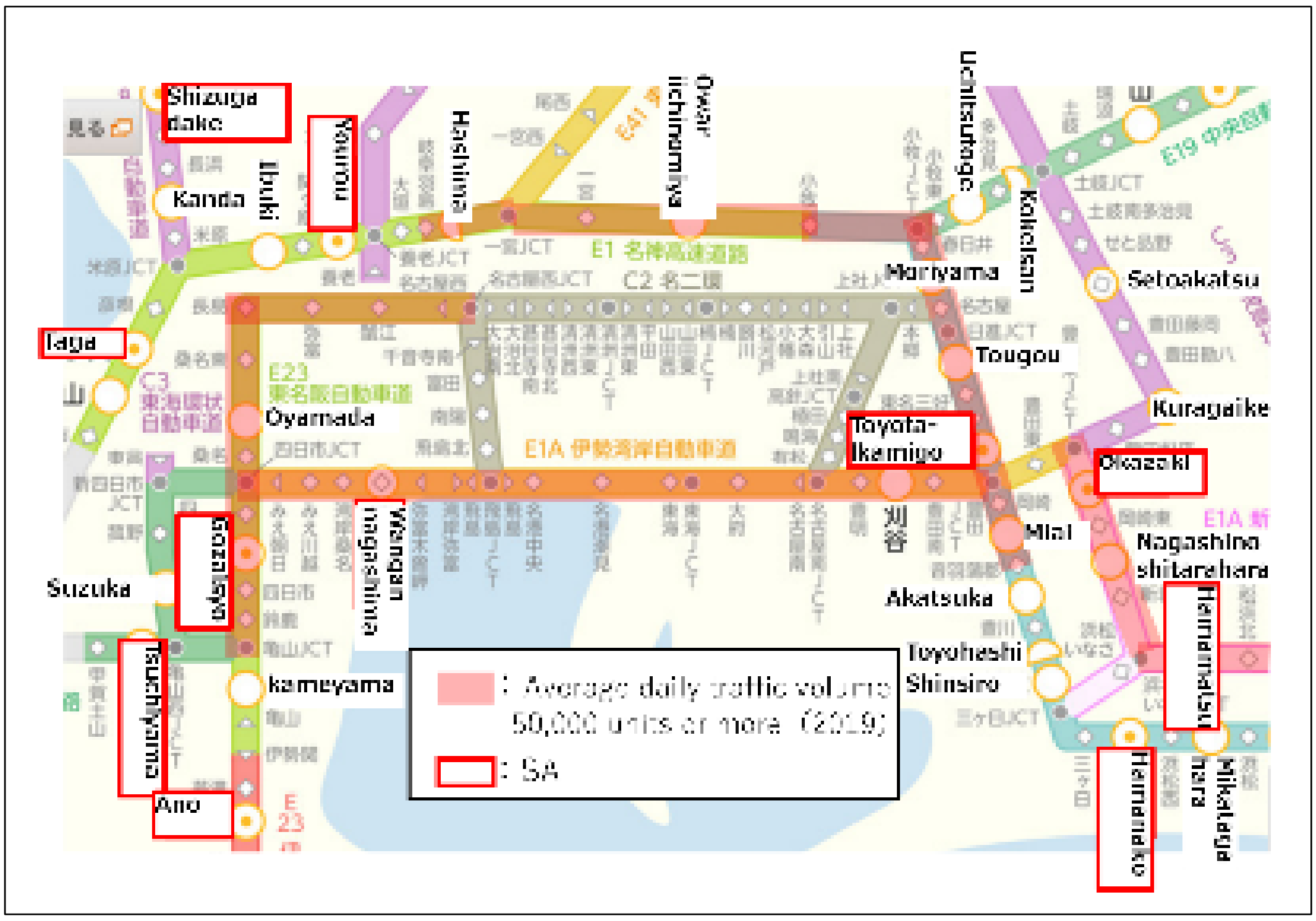

【Figure 9：Heavy Traffic Sections in Chubu Area】

The sample data is categorized by SAPA unit and year for heavy traffic and non-heavy traffic sections before analysis. SAPAs with a sample size of less than 40 during the period under study were excluded from the analysis because of the difficulty of the analysis. The following are the SAPAs for heavy traffic and non-heavy traffic sections.

| Road name | SAPA name |
|---|---|
| New Tomei Expressway | Surugawan-numazu (going away from Tokyo) |
| | Surugawan-numazu (going to Tokyo) |
| | Shimizu (aggregating) |
| | Hamamatsu (going away from Tokyo) |
| | Hamamatsu (going to Tokyo) |
| | Okazaki (aggregating) |
| Tomei Expressway | Kouhoku (going away from Tokyo) |
| | Kouhoku (going to Tokyo) |
| | Nakai (going away from Tokyo) |
| Tomei Expressway | Ayusawa (going to Tokyo) |
| | Ebina (going away from Tokyo) |
| | Ebina (going to Tokyo) |
| | Ashigara (going away from Tokyo) |
| | Ashigara (going to Tokyo) |
| | Toyotakamigou (going to Tokyo) |
| Chuo Expresswa | Dangouzawa (going away from Tokyo) |
| | Dangouzaka (going to Tokyo) |
| Higashimeihan expressway | Gozaisho (going to Tokyo) |

【Table 9: List of SAPAs in Heavy Traffic Sections】

| Road name | SAPA name |
|---|---|
| Tomei Expressway | Komakado (going away from Tokyo) |
| | Makinohara (going away from Tokyo) |
| | Makinohara (going to Tokyo) |
| | Hamanako(aggregating) |
| New Tomei Expressway | Shizuoka (going away from Tokyo) |
| | Shizuoka (going to Tokyo) |
| | Ensyumorimacchi (going away from Tokyo) |
| Ken-O Expressway | Atsugi (going away from Tokyo) |
| | Atsugi (going to Tokyo) |
| Odawara atsugi road | Oiso (going to Tokyo) |
| Chuo Expresswa | Ebina (going to Tokyo) |
| | Futaba (going away from Tokyo) |
| | Futaba (going to Tokyo) |
| Chuo Expresswa | Suwako (going away from Tokyo) |
| | Suwako (going to Tokyo) |
| Tokai-Hokuriku Expressway | Nagaragawa (going away from Tokyo) |
| Hokuriku Expressway | Najyo (going away from Tokyo) |
| | Nanjyo (going to Tokyo) |
| | Amagozen (going to Tokyo) |
| Meishin Expressway | Yourou (going away from Tokyo) |
| | Yourou (going to Tokyo) |
| | Taga (going away from Tokyo) |
| | Taga (going to Tokyo) |
| New Meishin Expressway Road | Suzuka (aggregating) |

【Table 10: List of SAPAs in Non-Heavy Traffic Sections】

## 5-1. Status for each fiscal year

### 5-1-1. FY 2018

The interest trends of SAPA users of heavy and non-heavy traffic sections in FY 2018 are as follows

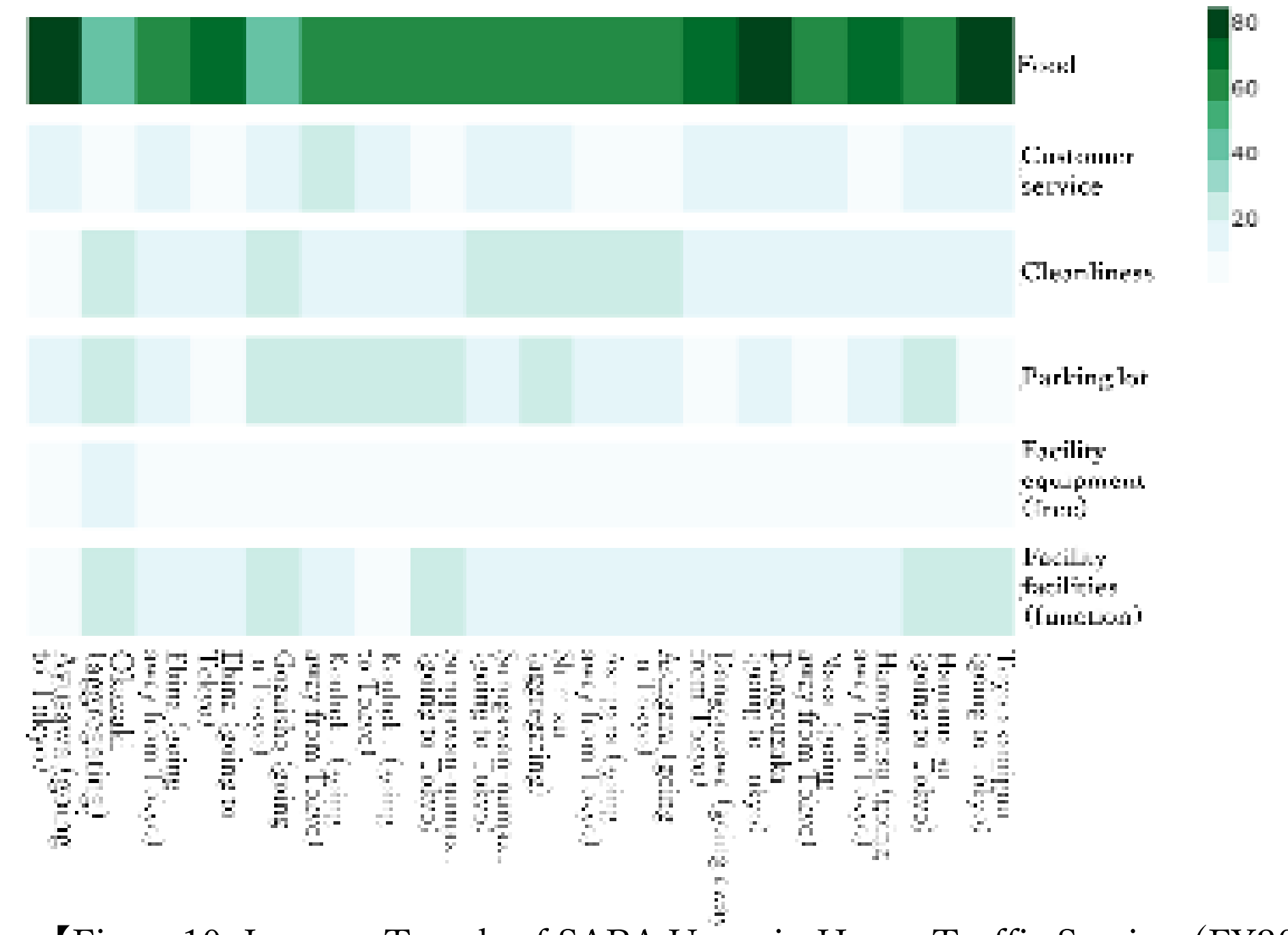

【Figure10: Interest Trends of SAPA Users in Heavy Traffic Section,(FY2018)】

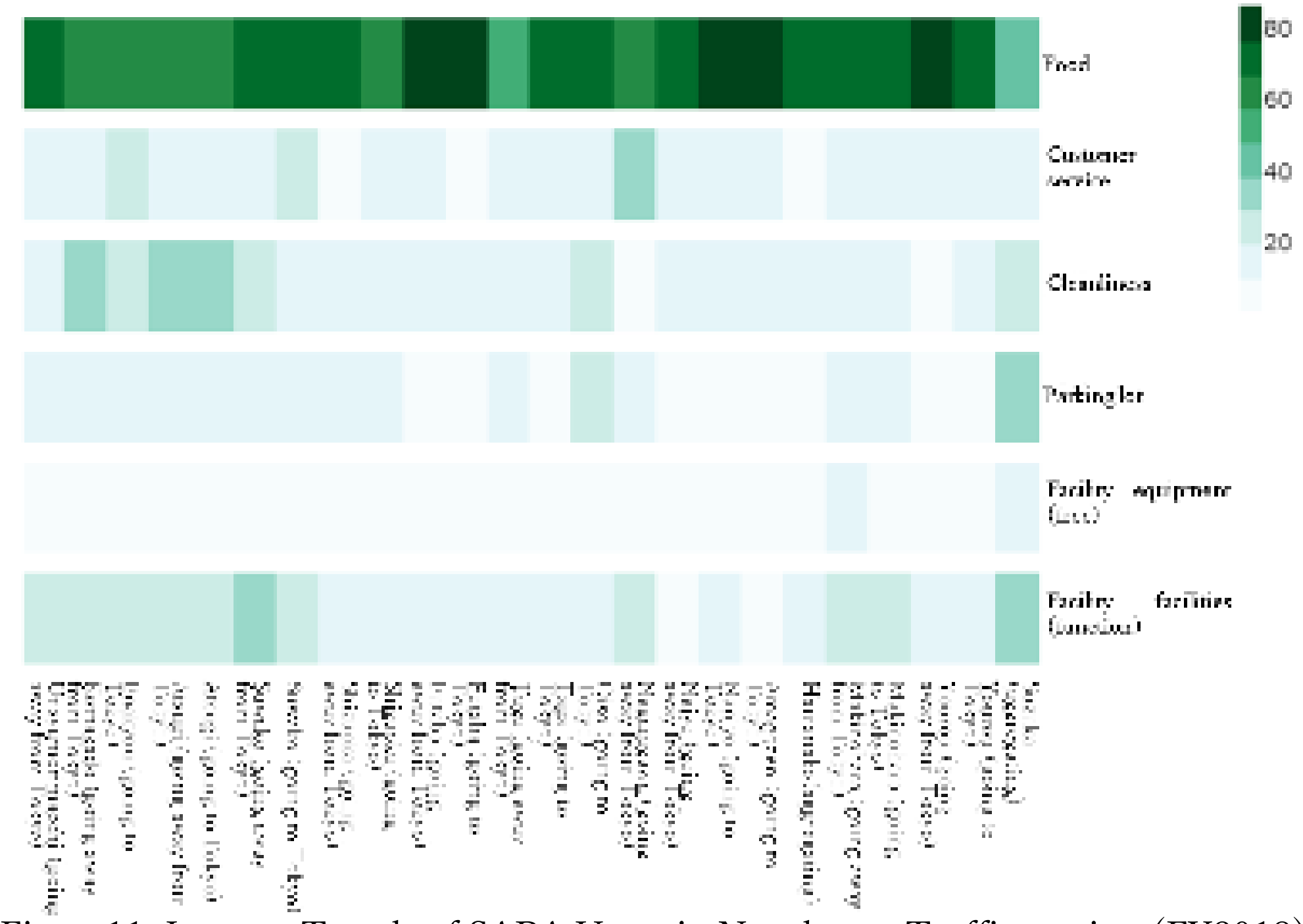

【Figure11: Interest Trends of SAPA Users in Non-heavy Traffic section,(FY2018)】

In the above graph, the darkest green squares indicate the items with the highest level of concern, and the next dark green squares indicate the items with the next highest level of concern. As a result, there was no difference between the SAPAs in the heavy traffic and non-heavy traffic sections in terms of the highest level of interest in meals. Therefore, to confirm the characteristics of the interests other than meals in each SAPA, we focused on the next item of interest. From this point of view, the following differences were found.

| | | Customer service | Cleanliness | Parking lots | Facility equipment (free) | Facility equipment (function) |
|---|---|---|---|---|---|---|
| Heavy traffic area | Number of SAPA | 4 | 7 | 7 | 0 | 2 |
| | Percentage | 20.0% | 35.0% | 35.0% | 0.0% | 10.0% |
| Non-heavy traffic areas | Number of SAPA | 7 | 11 | 1 | 0 | 8 |
| | Percentage | 25.9% | 40.7% | 3.7% | 0.0% | 29.6% |

【Table 11: Items of Next-level Interest, Aggregated Version (FY2018)】

Looking at the status of each, there was a difference in that in the heavy traffic section, opinions were more on the functional behavior side, such as parking lots, while in the non-heavy traffic section, opinions were more on the facility equipment (free) and free behavior side. In addition, there was a high level of interest in cleanliness on both sides.

Looking at the status of each, there was a difference in that in the heavy traffic section, opinions were more on the functional behavior side, such as parking, while in the non-heavy traffic section, opinions were more on the facility equipment (free) and free behavior side. In addition, there was a high level of interest in cleanliness on both sides.

The appearance of the opinions by the multidimensional scaling method is as follows.

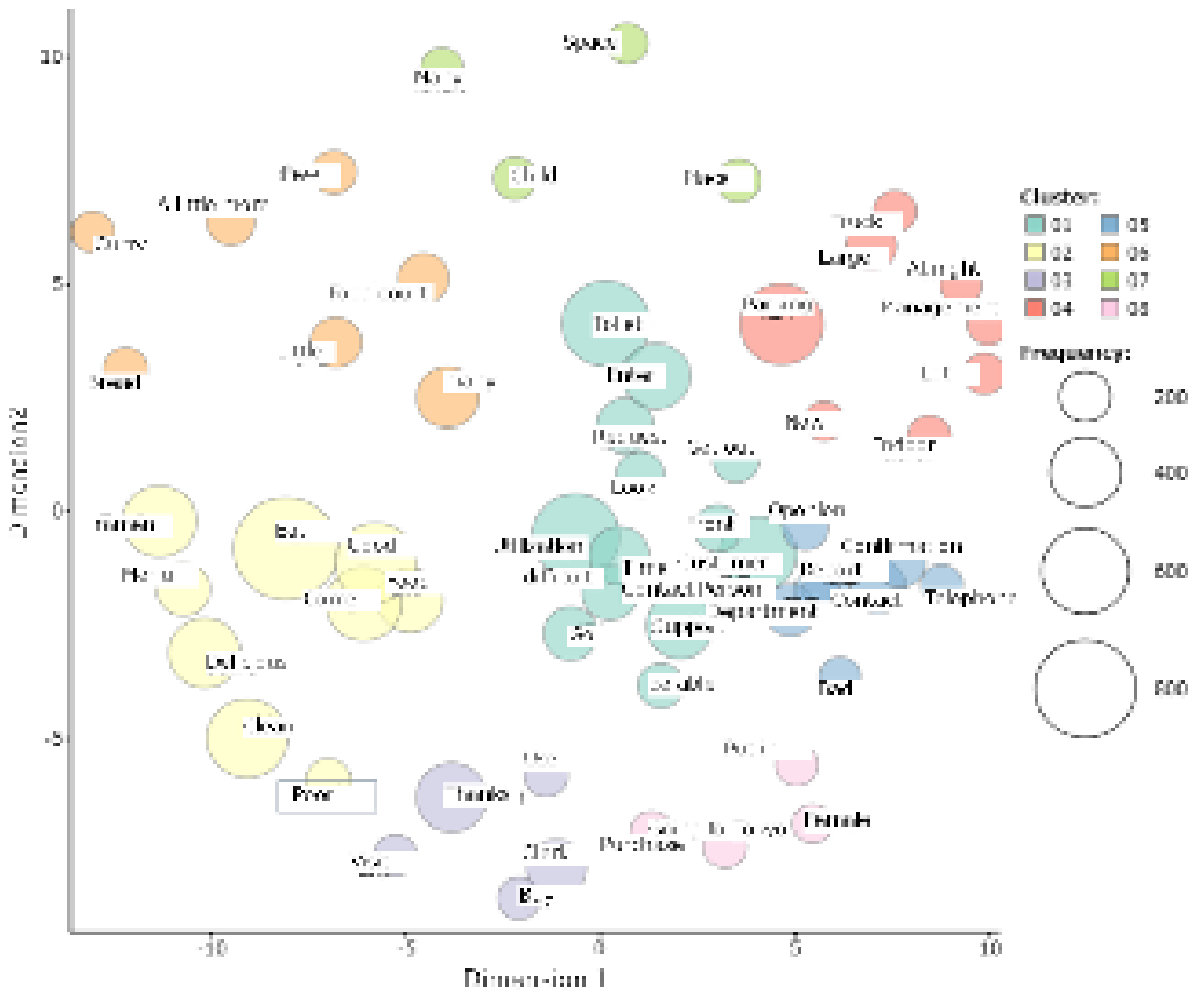

**The Setting of multi-dimensional scaling method analysis**
Word aggregation unit: H5
Vocabulary selection:
Selecting the minimum number of occurrences 120
Method：kruskal
Distance：Jaccard

【Figure 12: Appearance of Opinions in Heavy Traffic Section】

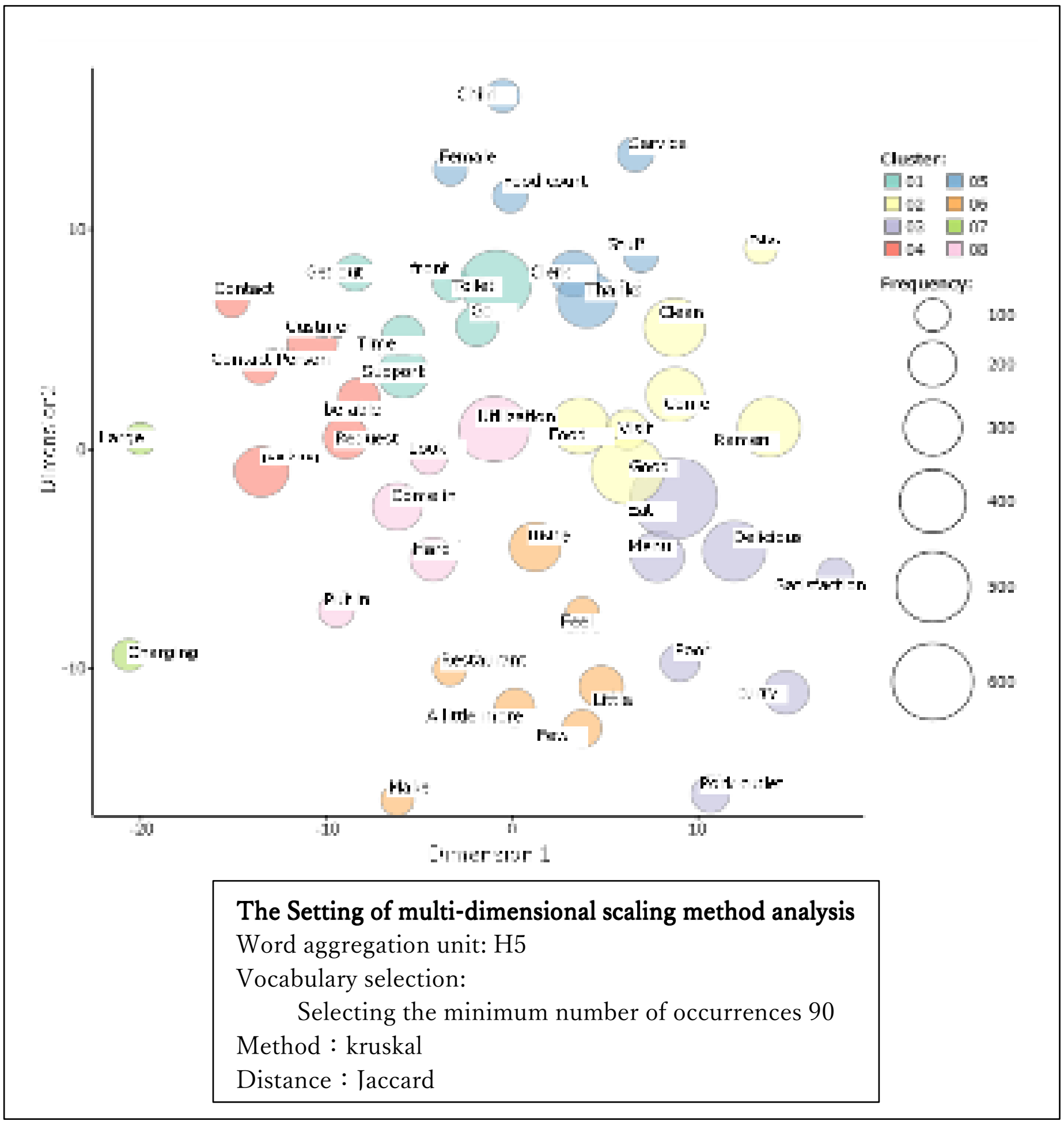

【Figure 13: Appearance of Opinions in Non-heavy Traffic Section】

According to the results, there were many opinions on meals that corresponded to functional behavior for both heavy traffic and non-heavy traffic sections. There were also many comments on parking in the heavy traffic section. On the other hand, for free activities, many comments were received for facilities and equipment (free) in the non-heavy traffic section. Next, we will check the evaluation status in each category.

To clarify whether each interest was favorable or unfavorable, the following categories related to emotional expression were added, to identify areas for improvement by looking at the degree to which the needs of each category were met and to make recommendations.

| Category | Target words |
|---|---|
| Good impression | Good, thankful, cheerful, comfortable, pleasant, etc. |
| bad impression | Noisy, slow, no good, terrible, uncomfortable, etc. |

【Table 12: Summary of Categories Related to Emotional Expression】

| Category | FY2018 | | FY2019 | | FY2020 | |
|---|---|---|---|---|---|---|
| | Good impression | Bad impression | Good impression | Bad impression | Good impression | Bad impression |
| Food | 0.704 | 0.102 | 0.615 | 0.093 | 0.516 | 0.115 |
| Customer service | – | – | 0.187 | 0.119 | – | – |
| Cleanliness | 0.165 | 0.113 | 0.189 | 0.106 | 0.173 | 0.118 |
| Parking lot | 0.103 | 0.124 | 0.23 | 0.128 | 0.113 | 0.167 |
| Facility equipment (free) | 0.227 | 0.14 | – | – | 0.095 | 0.042 |

【Table 13: Categories of Interest and Emotional Expressions】

In terms of meals, the degree of positive sentiment was high, indicating that the meals were well-received. On the other hand, the degree of negative feelings about the parking lot was higher than that of positive feelings, suggesting that users have a high demand for improvement.

| Category | Heavy traffic area | | Non-heavy traffic areas | |
|---|---|---|---|---|
| | Good impression | Bad impression | Good impression | Bad impression |
| Food | 0.702 | 0.098 | 0.724 | 0.094 |
| Customer service | 0.099 | 0.109 | 0.136 | 0.091 |
| Cleanliness | 0.147 | 0.112 | 0.161 | 0.121 |
| Parking lot | 0.084 | 0.124 | 0.074 | 0.097 |
| Facility equipment (free) | 0.131 | 0.134 | 0.14 | 0.118 |
| Facility equipment (function) | 0.023 | 0.057 | 0.017 | 0.046 |

【Table 14: Interest and Favorable and Unfavorable Feelings (FY2018)】

In both the heavy traffic and non-heavy traffic sections, more than 70% of the respondents had a favorable feeling toward food, indicating a high evaluation. For the other categories, there was no significant difference in the percentage of positive and negative feelings. However, there was a slight increase in the number of negative feelings about the parking lot in the heavy traffic section.

### 5-1-2. FY 2019

User interest trends for heavy and non-heavy traffic sections for FY 2019 are as follows

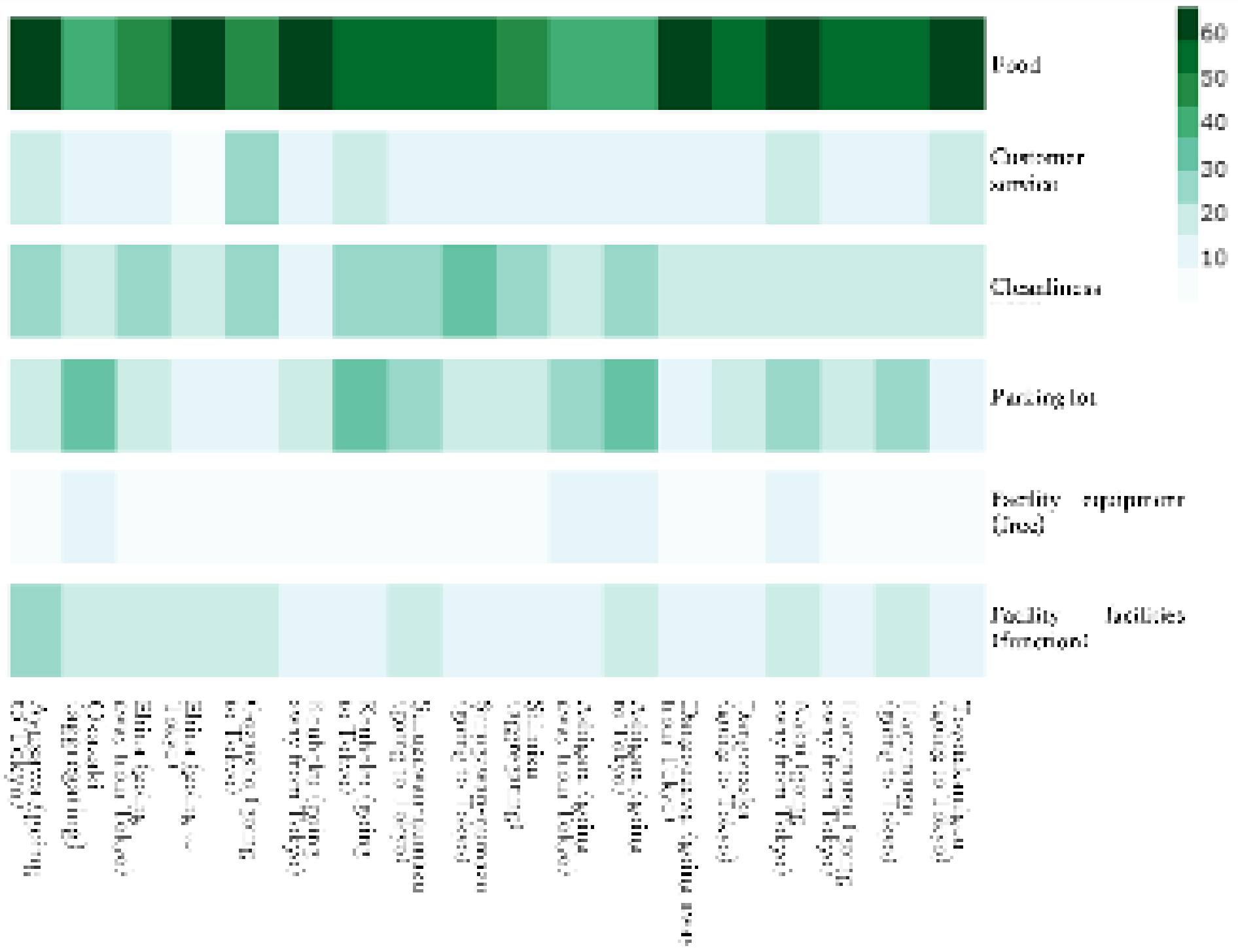

【Figure14: Interest Trends of SAPA Users in Heavy Traffic Section (FY2019)】

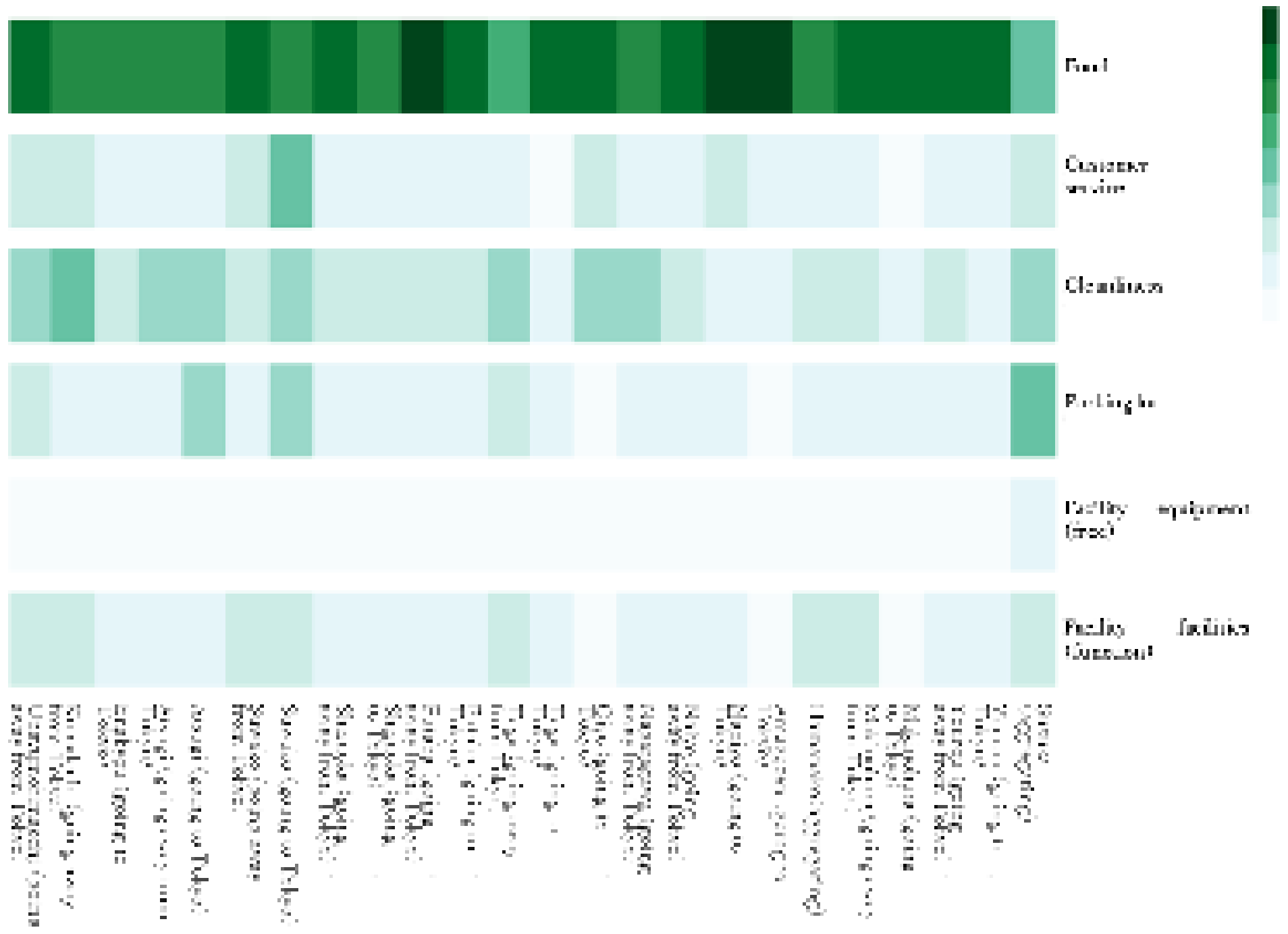

【Figure15: Interest Trends of SAPA Users in Non-heavy Traffic Section (FY2019)】

In FY 2019, as in the previous year, meals were the most common item in both heavy and non-heavy traffic sections. The runner-up items are as follows.

| | | Customer service | Cleanliness | Parking lots | Facility equipment (free) | Facility equipment (function) |
|---|---|---|---|---|---|---|
| Heavy traffic area | Number of SAPA | 0 | 9 | 9 | 0 | 0 |
| | Percentage | 0.0% | 50.0% | 50.0% | 0.0% | 0.0% |
| Non-heavy traffic areas | Number of SAPA | 3 | 19 | 1 | 0 | 2 |
| | Percentage | 12.0% | 76.0% | 4.0% | 0.0% | 8.0% |

【Table 15: Items of Interest for Secondary Points, Aggregated Version (FY2019)】

The results show that the heavy traffic sections are more concerned with cleanliness and parking lots, while the non-heavy traffic sections are more concerned with cleanliness.

Next, the appearance of opinions using the multidimensional scaling method is shown below.

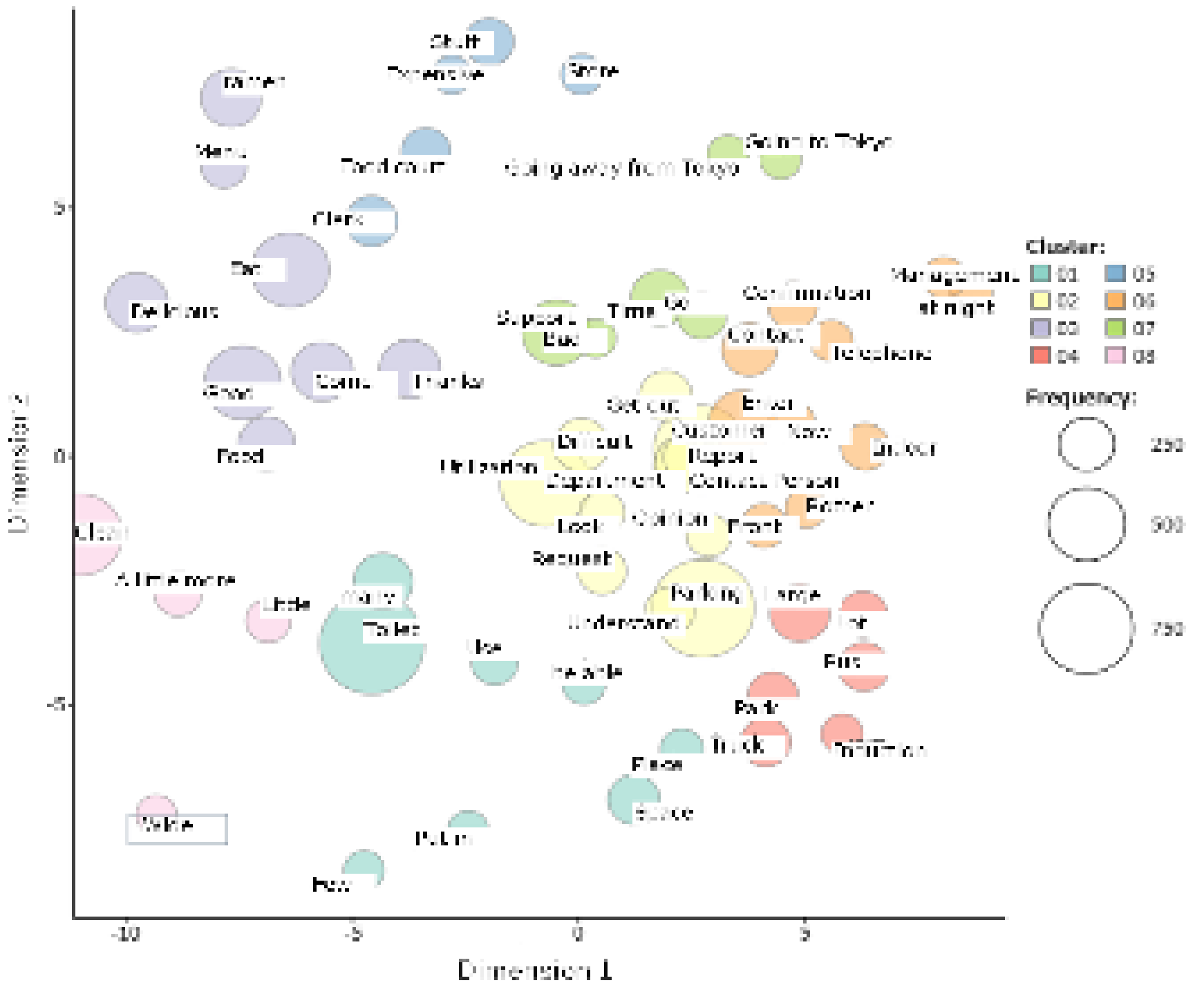

**The Setting of multi-dimensional scaling method analysis**
Word aggregation unit: H5
Vocabulary selection:
Selecting the minimum number of occurrences 120
Method：kruskal
Distance：Jaccard

【Figure 16: Appearance of Opinions in Heavy Traffic Section】

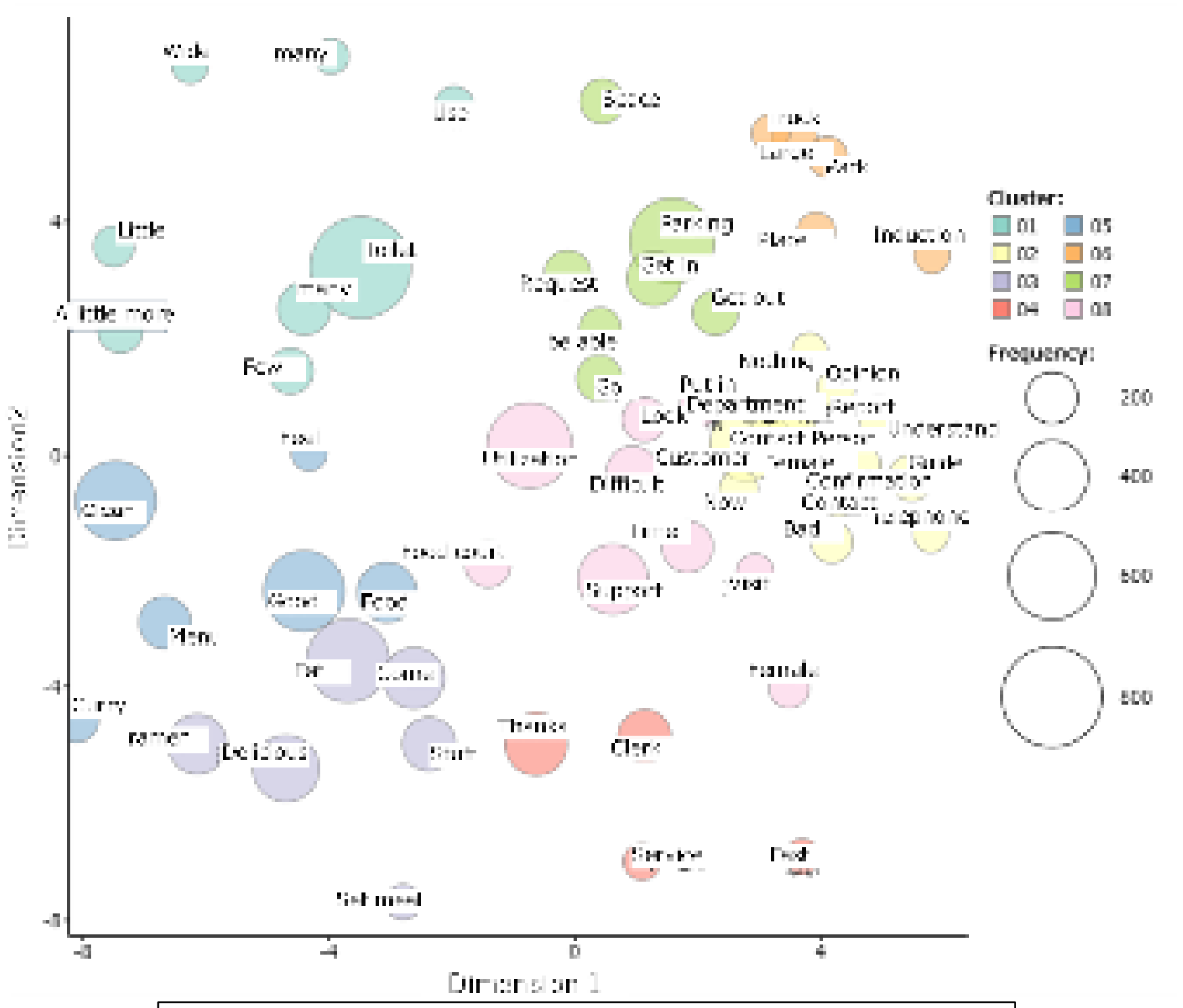

**The Setting of multi-dimensional scaling method analysis**
Word aggregation unit: H5
Vocabulary selection:
Selecting the minimum number of occurrences 90
Method：kruskal
Distance：Jaccard
Number of clusters setting：8

【Figure 17: Appearance of Opinions in Non-heavy Traffic Section】

In both the heavy traffic and non-heavy traffic sections, there were many opinions on meals, a functional behavior like the previous year. In addition, there were many opinions about parking in the heavy traffic section.

As for the free time activities, there were many opinions about customer service in the non-heavy traffic section.

Next, the similarity between the positive and negative feelings is as follows.

| Category | Heavy traffic area | | Non-heavy traffic areas | |
|---|---|---|---|---|
| | Good impression | Bad impression | Good impression | Bad impression |
| Food | 0.587 | 0.092 | 0.649 | 0.094 |
| Customer service | 0.125 | 0.091 | 0.154 | 0.102 |
| Cleanliness | 0.181 | 0.095 | 0.181 | 0.108 |
| Parking lot | 0.098 | 0.135 | 0.094 | 0.116 |
| Facility equipment (free) | 0.138 | 0.117 | 0.137 | 0.129 |
| Facility equipment (function) | 0.024 | 0.06 | 0.023 | 0.039 |

【Table 16: Interest and Favorable and Unfavorable Feelings (FY2019)】

Both heavy-traffic and non-heavy-traffic sections received high ratings for food, but the degree of positive sentiment has declined since the previous year. As for the other categories, bad feelings about parking are relatively high in the heavy traffic section.

### 5-1-3. FY 2020

The following are the trends of interest among SAPA users of heavy and non-heavy traffic sections in FY2020.

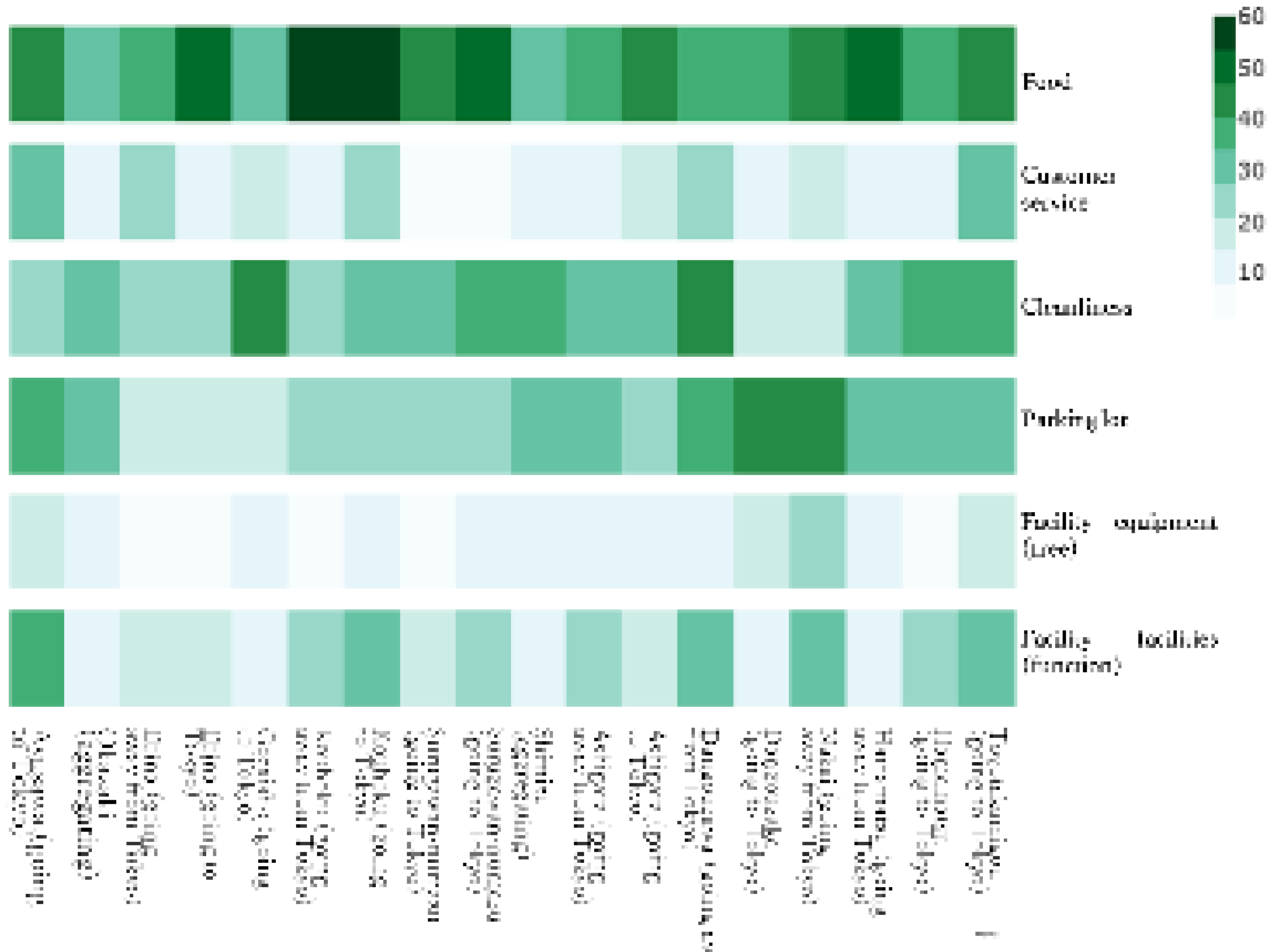

【Figure18: Interest Trends of SAPA Users in Heavy Traffic Section (FY2020)】

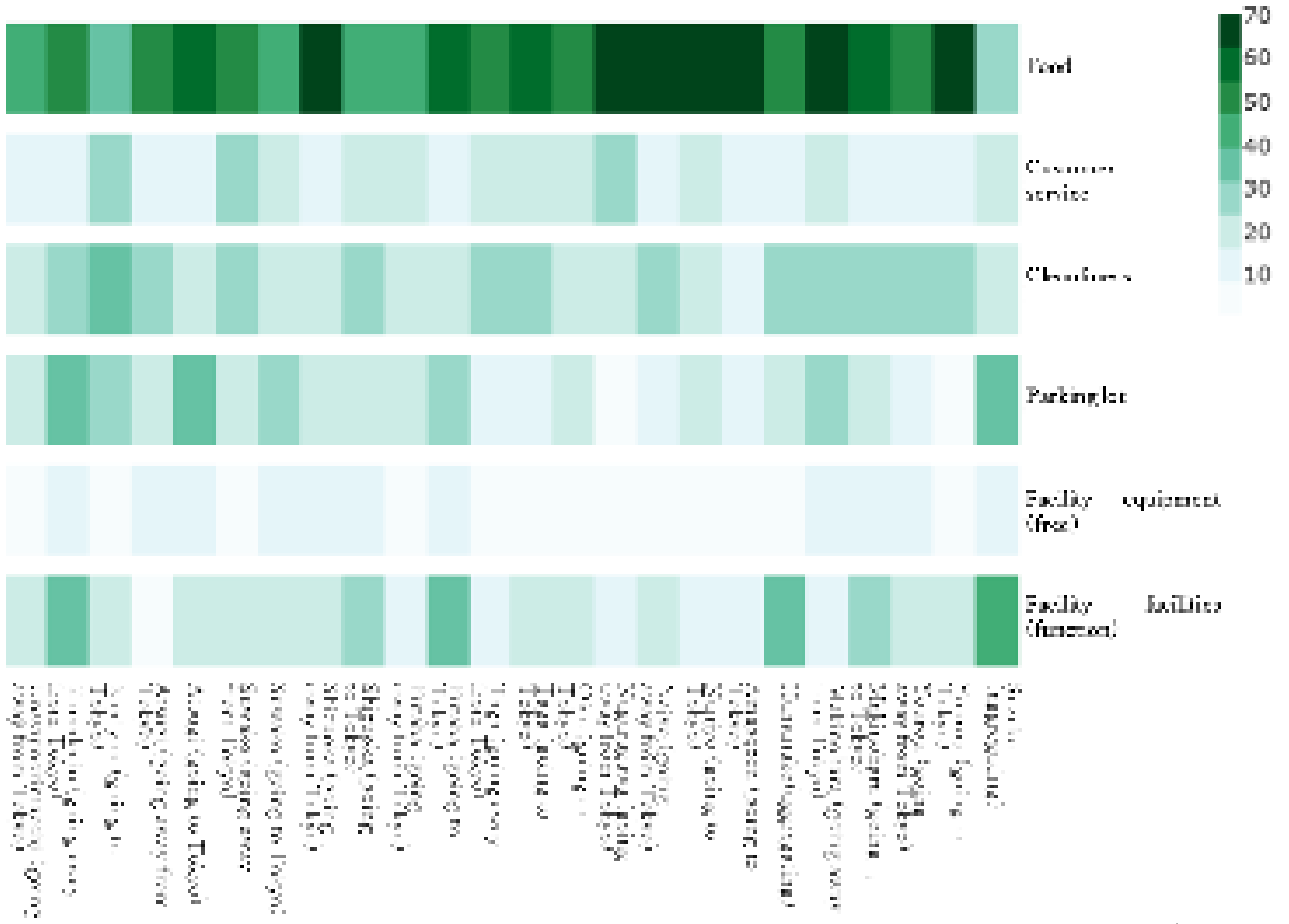

【Figure19: Interest Trends of SAPA Users in Non-heavy Traffic Section (FY2020)】

In both heavy traffic and non-heavy traffic sections, food is still the most common item, but in some SAPAs in heavy traffic sections, cleanliness and parking lots are the most common items, indicating a change.

The status of the next highest scores is as follows.

| | | Customer service | Cleanliness | Parking lots | Facility equipment (free) | Facility equipment (function) | Customer service |
|---|---|---|---|---|---|---|---|
| Heavy traffic area | Number of SAPA | 3 | 0 | 8 | 6 | 0 | 2 |
| | Percentage | 15.8% | 0.0% | 42.1% | 31.6% | 0.0% | 10.5% |
| Non-heavy traffic areas | Number of SAPA | 0 | 4 | 15 | 8 | 0 | 1 |
| | Percentage | 0.0% | 14.3% | 53.6% | 28.6% | 0.0% | 3.6% |

【Table 17: Items of Next-level Interest Aggregated Version (FY2020)】

In both heavy traffic and non-heavy traffic sections, there is interest in cleanliness. In addition, in the non-heavy traffic section, there was a new interest in parking lots, indicating a change from the previous year.

Next, the appearance of opinions using the multidimensional scaling method is shown below.

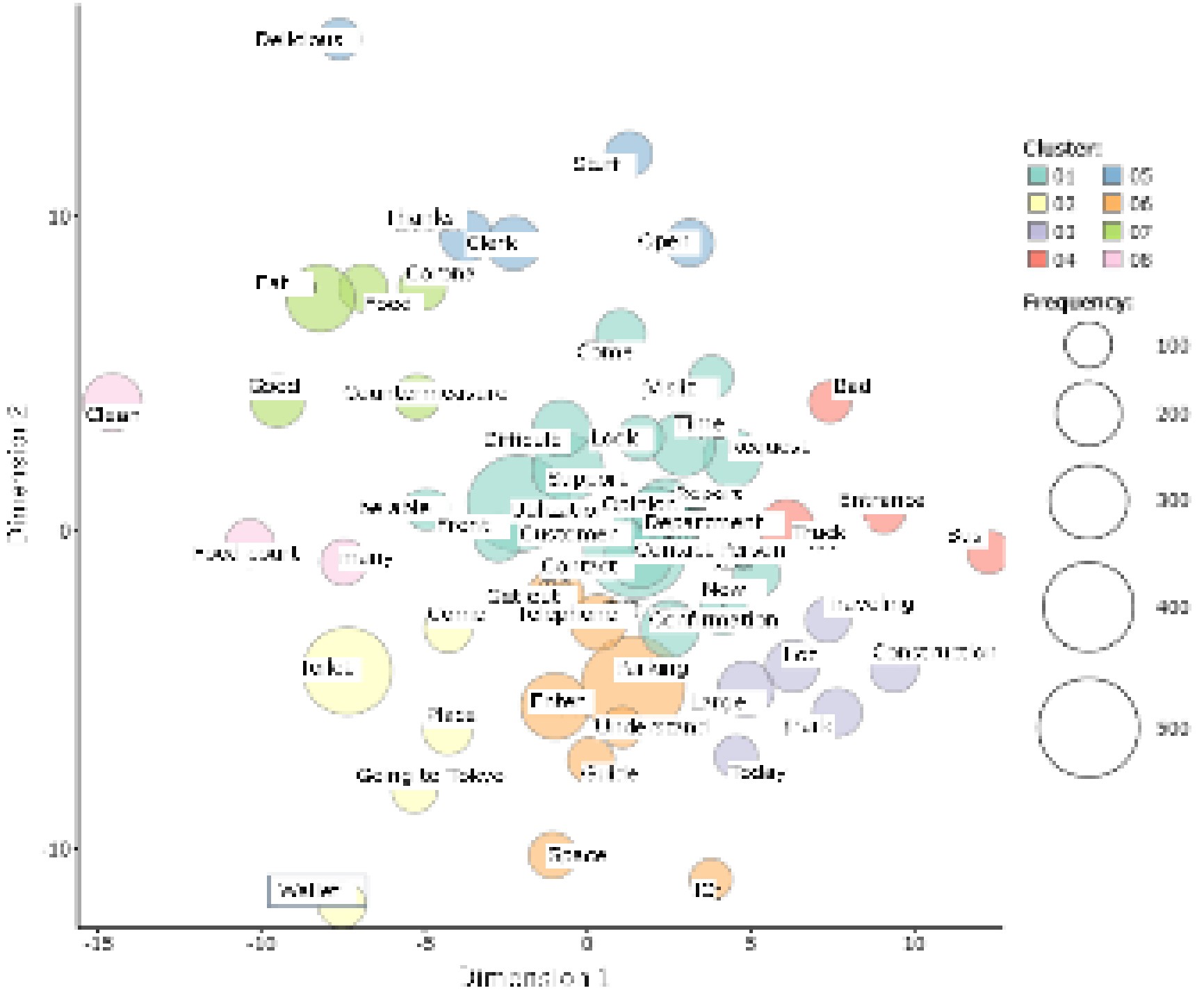

**The Setting of multi-dimensional scaling method analysis**
Word aggregation unit: H5
Vocabulary selection:
Selecting the minimum number of occurrences 80
Method：kruskal
Distance：Jaccard
Number of clusters setting：8

【Figure 20: Appearance of Opinions in Heavy Traffic Section】

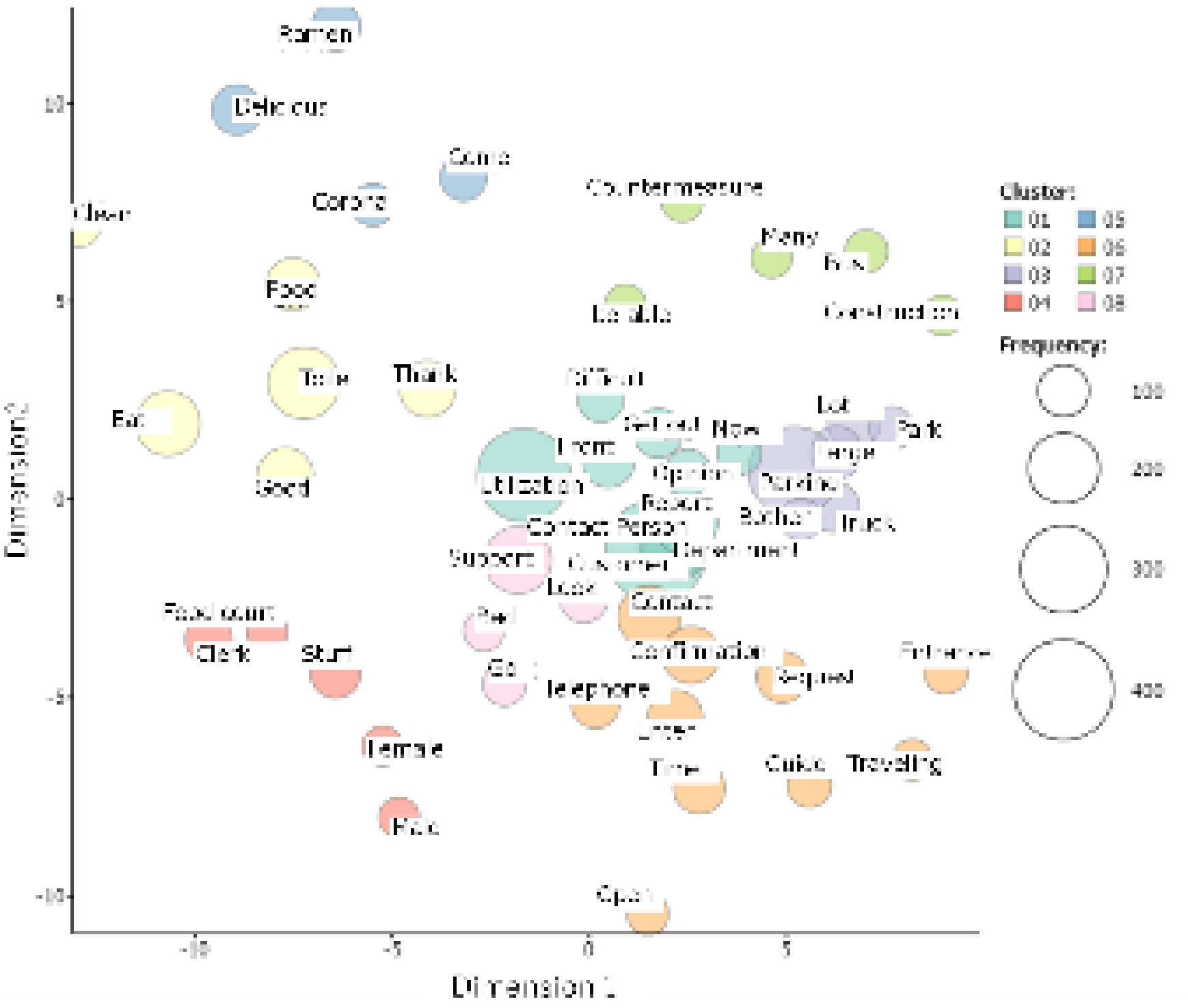

**The Setting of multi-dimensional scaling method analysis**
Word aggregation unit: H5
Vocabulary selection:
Selecting the minimum number of occurrences 60
Method：kruskal
Distance：Jaccard
Number of clusters setting：8

【Figure 21: Appearance of Opinions in Non-heavy Traffic Section】

In FY2020, there is a change in the content of opinions regarding functional behavior. In both the heavy-traffic and non-heavy-traffic sections, eating was the most popular item, but there was also a lot of interest in parking. In terms of free activities, opinions on facilities and equipment (free) increased in the heavy traffic section. In addition, reflecting the Corona disaster, the word "corona" was newly mentioned frequently in both heavy and non-heavy traffic sections.

The similarity matrix between positive and negative emotions is shown below.

| Category | Heavy traffic area | | Non-heavy traffic areas | |
|---|---|---|---|---|
| | Good impression | Bad impression | Good impression | Bad impression |
| Food | 0.49 | 0.102 | 0.595 | 0.123 |
| Customer service | 0.144 | 0.144 | 0.164 | 0.135 |
| Cleanliness | 0.206 | 0.125 | 0.173 | 0.16 |
| Parking lot | 0.119 | 0.162 | 0.091 | 0.156 |
| Facility equipment (free) | 0.139 | 0.113 | 0.13 | 0.145 |
| Facility equipment (function) | 0.051 | 0.108 | 0.032 | 0.082 |

【Table 18: Interest and Favorable and Unfavorable Feelings (FY2020)】

In FY2020, we continue to receive a certain level of evaluation for food, but the degree of evaluation continues to decline. For the others, there were no significant changes from the previous year.

## 5-2. Differences in needs between SAPA in heavy traffic sections and SAPA in non-heavy traffic sections

The results of the analysis for each year can be summarized as follows.

| | | Heavy traffic area | Non-heavy traffic areas |
|---|---|---|---|
| Free Movement | | Distribution and frequency High | Distribution and frequency High |
| | Customer service | Interest Medium | Interest High |
| | Facility equipment (free) | Interest Low | Interest High |
| Functional Movement | | Distribution and frequency High | Distribution and frequency Low |
| | Food | Interest High | Interest High |
| | Parking lots | Interest High | Interest Low |
| | Facility equipment (function) | Interest Low | Interest Low |
| Cleanliness | | Interest High | Interest High |

【Table 19: Comparison of SAPA Users' interests (FY2018)】

| | | Heavy traffic area | Non-heavy traffic areas |
|---|---|---|---|
| Free Movement | | Distribution and frequency High | Distribution and frequency High |
| | Customer service | Interest Low | Interest Medium |
| | Facility equipment (free) | Interest Low | Interest Low |
| Functional Movement | | Distribution and frequency High | Distribution and frequency Low |
| | Food | Interest High | Interest High |
| | Parking lots | Interest High | Interest Low |
| | Facility equipment (function) | Interest Low | Interest Low |
| Cleanliness | | Interest High | Interest High |

【Table 20: Comparison of SAPA Users' interests (FY2019)】

| | | Heavy traffic area | Non-heavy traffic areas |
|---|---|---|---|
| Free Movement | | Distribution and frequency High | Distribution and frequency High |
| | Customer service | Interest Low | Interest Low |
| | Facility equipment (free) | Interest Medium | Interest Low |
| Functional Movement | | Distribution and frequency High | Distribution and frequency High |
| | Food | Interest High | Interest High |
| | Parking lots | Interest High | Interest High |
| | Facility equipment (function) | Interest Low | Interest Low |
| Cleanliness | | Interest High | Interest High |

【Table 21: Comparison of SAPA Users' interests (FY2020)】

In terms of functional behavior, food was the highest in most SAPAs throughout the period under study. This was followed by a high interest in parking lots, especially in the heavy traffic sections. This situation will be seen in the non-heavy traffic sections in FY2020.

As for free behavior, the non-heavy traffic sections have a high interest in this area. In FY2020, some SAPAs in the heavy traffic section also showed interest in this area, but the percentage was not large. As for cleanliness, both sections have high interest in it during the period covered.

As for the individual opinions included in the functional behavior, those related to meals were the most frequent. Next in frequency were words related to parking lots. Words related to restrooms were also frequently used. As for free behavior, the frequency of those related to customer service was high.

In FY2020, among the functional behaviors in the non-heavy traffic section, there was an increase in interest in parking lots, which is like the interesting trend in the heavy traffic section. In addition, the word "Corona" was used in both heavy and non-heavy traffic sections, reflecting the covid-19. In both categories, the word "corona" was associated with items related to food, suggesting that customers were concerned about covid-19 countermeasures when using food courts. However, the frequency was not significant.

To see the factors behind this change, we use Futaba SA (going to Tokyo), one of the SAs in the non-heavy traffic section, as a sample to check the attributes of the customers who expressed their opinions.

| | | FY2018 | FY2019 | FY2020 |
|---|---|---|---|---|
| Female | Under 20s | 46 | 31 | 5 |
| | 20s～50s | 43 | 37 | 21 |
| | Over 60s | 24 | 9 | 6 |
| | Age unknown | 13 | 8 | 10 |
| | Subtotal | 126 | 85 | 42 |
| Male | Under 20s | 19 | 22 | 5 |
| | 20s～50s | 87 | 71 | 25 |
| | Over 60s | 39 | 29 | 14 |
| | Age unknown | 21 | 33 | 29 |
| | Subtotal | 166 | 155 | 73 |
| Gender unknown | | 188 | 188 | 28 |
| Total | | 480 | 480 | 268 |

【Table 22: Customer Comments Attribute Status in Futaba SA (Going to Tokyo)】

First, the number of opinions in FY2020 has decreased significantly compared to the previous year. Also, the ratio of women to men has reversed from the previous year, when most opinions were held by women. In addition, by age group, the number of opinions from those in their twenties or younger has decreased significantly.

## 6. Conclusions and recommendations

Regarding the first hypothesis, the difference in needs for SAPA by route, most of the routes were highly interested in food, because food was the most common trend in the overall sample data. However, some routes showed interest in other options of the functional behavior, for example, parking lots. These routes, like the Shin-Meishin expressway, are used for east-west travel and may reflect the high ratio of trucks. This was observed for some routes.

The second hypothesis is that there is a difference in needs for SAPA due to traffic volume, Customers are the most interested in food in both sections. But when referring to the runner-up interest, it revealed that the difference in traffic volume affected the difference in needs, like parking lots or facility equipment (free).

Regarding the third hypothesis, customer changes need before and after the Covid-19 disaster, there were no major changes in the situation in terms of route and traffic volume before and after FY2020, even for cleanliness, which is assumed to be related to Covid-19. However, there was an increase in interest in parking lots in non-heavy traffic areas. In conclusion, the changes in customer needs before and after the Covid-19 disaster were less changed.

About previous studies, Umayahara et al. (2017) found that SAPA users engaged in functional and free behavior to about the same extent, but in this study, most opinions were related to functional behavior 'food'. It revealed that SAPA users perform both types of behaviors almost equally, but in terms of interest, they are more interested in functional behavior. So, focusing on the functional behavior aspect of SAPA improvement is effective in gaining users' satisfaction with SAPA.

Nishii et al. (2017) and Shiino et al. (2011), some points that were not listed as standing factors were found in the opinions. In addition, there were some differences in the frequency of opinions of each standing factor. This suggests that there is a difference between the drop-in factors and the needs of the users when they visit SAPA.

In addition, Nishii et al. (2017) found that Class 2, who use SAPA for sightseeing and leisure purposes, cited food as a stopping factor. They may visit a wide range of SAPA because most opinions from customers are about food. On the other hand, there were some routes where opinions other than food were frequently expressed, such as parking lot and facilities (free). This matter may indicate the possibility of the existence of classes other than Class 2.

If the above perspective is to be utilized in the future development of SAPA and the expansion of its services, the following measures can be considered. First, in the planning of SAPA renovation, the characteristics of the opinions in each route or each traffic volume situation will help us formulate a maintenance plan that better meets the needs of SAPA users. Based on these characteristics, it is possible to formulate a maintenance plan that is more suitable for the SAPA user's group.

It is also possible to use this trend of users' interest to differentiate SAPA. The fact that there are many opinions about food on most routes can be used. For example, to resolve parking lot congestion, reviewing the food service encourages the division of SAPAs into those for business use and those for sightseeing and leisure use. It may be to promote the voluntary selection of SAPAs by users. This is a new way to address the shortage of parking lots, adding to expanding the parking lots and using road information boards.

Concerning the effect of the covid-19, a change in interest was observed in the non-heavy traffic section in this study. It is necessary to observe whether this change is brought by whether a temporary change in the SAPA user's group or a change in the SAPA user behavior itself due to the covid-19.

The first issue in this research is the bias of the sample data. Since some parts of SAPA users post their opinions, the data may not represent the tendency of all SAPA users. It is necessary to compensate for this bias by collecting information from other media such as SNS.

The second is the collection of samples after the covid-19. Due to the decrease in traffic caused by covid-19, the demographics of the SAPA users changed in 2020. To reveal whether the demographics of the users or the behavior of users makes the change of interesting of SAPAs opinions especially in non-heavy traffic section, the data in recovering traffic volume after ceasing this disaster are needed.

## Acknowledgments

I would like to express my gratitude to my advisors, Professor Takuji Takemoto, Professor Yoshimi Kawamoto and Assistant Professor Sachiyo Kawakami for their enthusiastic guidance throughout my research.